\documentclass[onecolumn,british,english,twocolumn]{IEEEtran}
\usepackage[T1]{fontenc}
\usepackage[latin9]{inputenc}
\usepackage[active]{srcltx}
\usepackage{array}
\usepackage{cprotect}
\usepackage{float}
\usepackage{bbding}
\usepackage{multirow}
\usepackage{varwidth}
\usepackage{amsmath}
\usepackage{amsthm}
\usepackage{amssymb}
\usepackage{graphicx}

\makeatletter

\providecommand{\tabularnewline}{\\}
\newenvironment{cellvarwidth}[1][t]
    {\begin{varwidth}[#1]{\linewidth}}
    {\@finalstrut\@arstrutbox\end{varwidth}}
\floatstyle{ruled}
\newfloat{algorithm}{tbp}{loa}
\providecommand{\algorithmname}{Algorithm}
\floatname{algorithm}{\protect\algorithmname}

\theoremstyle{plain}
\newtheorem{thm}{\protect\theoremname}
\theoremstyle{definition}
\newtheorem{example}[thm]{\protect\examplename}
\theoremstyle{plain}
\newtheorem{lem}[thm]{\protect\lemmaname}
\theoremstyle{plain}
\newtheorem{prop}[thm]{\protect\propositionname}

\usepackage{cite} 
\usepackage[margin=8pt,font=footnotesize]{caption}
\usepackage{algorithm}
\usepackage{algpseudocode}

\usepackage{amsmath}  
\allowdisplaybreaks

\makeatother

\usepackage{babel}
\addto\captionsbritish{\renewcommand{\algorithmname}{Algorithm}}
\addto\captionsbritish{\renewcommand{\examplename}{Example}}
\addto\captionsbritish{\renewcommand{\lemmaname}{Lemma}}
\addto\captionsbritish{\renewcommand{\propositionname}{Proposition}}
\addto\captionsbritish{\renewcommand{\theoremname}{Theorem}}
\addto\captionsenglish{\renewcommand{\examplename}{Example}}
\addto\captionsenglish{\renewcommand{\lemmaname}{Lemma}}
\addto\captionsenglish{\renewcommand{\propositionname}{Proposition}}
\addto\captionsenglish{\renewcommand{\theoremname}{Theorem}}
\providecommand{\examplename}{Example}
\providecommand{\lemmaname}{Lemma}
\providecommand{\propositionname}{Proposition}
\providecommand{\theoremname}{Theorem}

\begin{document}
\title{A Track-Before-Detect Trajectory Multi-Bernoulli Filter for Nested
Superpositional Measurements \thanks{This work was jointly supported by the University of Liverpool and
Leonardo UK.}}
\author{Sion Lynch\thanks{Sion Lynch and Lee Devlin are with the Department of Electrical Engineering
and Electronics, University of Liverpool, L69 3GJ U.K. (emails: \{sion.lynch,ljdevlin\}@liverpool.ac.uk).}, \'Angel F. Garc\'ia-Fern\'andez\thanks{\'Angel F. Garc\'ia-Fern\'andez is with the Information Processing
and Telecommunications Center, ETSI de Telecomunicaci\'on, Universidad
Polit\'ecnica de Madrid, 28040 Madrid, Spain (email: \mbox{angel.garcia.fernandez@upm.es}).} and Lee Devlin\thanks{This paper has supplementary downloadable material available at http://ieeexplore.ieee.org,
provided by the author. The material includes the appendices. This
material is 200KB in size.}}
\maketitle
\begin{abstract}
This paper proposes the Trajectory-Information Exchange Multi-Bernoulli
(T-IEMB) filter to estimate sets of alive and all trajectories in
track-before-detect applications with nested superpositional measurements.
This measurement model has a nested architecture with two layers.
The first layer includes superpositional hidden variables which are
then mapped in the second layer to the conditional mean and covariance
of the measurement, enabling it to model a broad range of measurement
models. This paper also presents a Gaussian implementation of the
T-IEMB filter, which performs the update by approximating the conditional
moments of the measurement model, a computationally light filtering
solution. Simulation results for a non-Gaussian radar-based tracking
scenario demonstrate the performance of two Gaussian T-IEMB implementations,
which provide improved tracking performance compared to state-of-the-art
particle filters for track-before-detect, at a reduced computational
cost.
\end{abstract}

\begin{IEEEkeywords}
Multi-target tracking, sets of trajectories, track-before-detect,
superpositional measurements, multi-Bernoulli.
\end{IEEEkeywords}

\section{Introduction}

Multi-target tracking (MTT) involves estimating the evolution of object
states using sensor information corrupted by noise \cite{Blackman1999}.
Common applications of target tracking include radar-based surveillance
\cite{Ristic2025}, autonomous surface vehicles \cite{Yang2020},
and space surveillance \cite{Delande2019}. Typical target tracking
schemes use point detections, which are extracted from raw sensor
measurements using a thresholding scheme, and subsequently used to
perform tracking. Track-before-detect (TkBD) \cite{Davies2024,Ristic2020a,Bosser2025,Liang2023,Wu2023,Yi2020,Zheng2024}
is an alternative approach in which sensor measurements forego any
thresholding, such that raw sensor measurements are used to perform
tracking. This is advantageous when tracking targets with a weak return
signal, as they are unlikely to produce a detection using traditional
processing or thresholding schemes.

Early approaches for TkBD typically assume there is at most a single
target present \cite{Kirubarajan1995}\cite{Salmond2001}. A natural
approach to tackle TkBD problems with an unknown and time-varying
number of targets is to use the Random Finite Set (RFS) \cite{Mahler2014}
framework. RFS-based filters model the multi-target state of interest
as a set, which captures uncertainty on both the number and states
of the targets. The RFS framework therefore provides a unified theoretical
framework for handling target birth, death and updating existing state
estimates.

Derivations of RFS-based filters for TkBD require a suitable choice
of measurement model, which characterises the probability distribution
of the received signal depending on the number of targets and their
states. The relevant properties of typical TkBD measurement models
in the literature are presented in Table \ref{tab:SensorModelProperties}.
We proceed to discuss the properties and applications of each listed
measurement model.

For modelling general TkBD sensors, one can use a general measurement
model as in \cite{Morelande2007,F.GarciaFernandez2013,GarciaFernandez2016,Li2018,Papi2015a}.
This model makes no assumption on the measurement density given the
set of targets i.e. Gaussian or non-Gaussian, and may not have any
superpositional properties. Using this model requires the use of Sequential
Monte Carlo (SMC) or particle filtering techniques to perform the
inference task. Additionally, it was shown in \cite{Yi2013} that
using an independent target approximation, as also used in \cite{Morelande2007,F.GarciaFernandez2013,GarciaFernandez2016},
is beneficial to address the curse of dimensionality in particle filtering.
For general measurement models, references \cite{GarciaFernandez2016,Li2018,Papi2015a}
make use of the Labelled RFS (LRFS) approach in which targets are
uniquely labelled upon birth, to sequentially estimate trajectories
from labelled target estimates.

Alternatively, the measurement model may be standard superpositional
\cite[Chapter 19]{Mahler2014}. Under this sensor model, the measurement
density given the set of targets is Gaussian with a conditional mean
that is superpositional on the contribution of each target, and a
fixed covariance matrix. Such measurements typically model the received
signal at a sensor, such as an acoustic \cite{Nannuru2013b} or active/passive
radio-frequency sensor \cite{Nannuru2013c}, where the received signal
originates from multiple emitters and contains additive noise \cite{Beaudeau2015}.

The standard superpositional property enables the development of a
variety of TkBD filters. For instance, a Probability Hypothesis Density
(PHD) filter and Cardinalised PHD (CPHD) filter for standard superpositional
measurements have been proposed, using an SMC \cite{Nannuru2013b},
or a Gaussian mixture \cite{Hauschildt2011} implementation. Multi-Bernoulli
(MB) filters have also been extended to standard superpositional measurements
\cite{Nannuru2013}, with an SMC implementation proposed in \cite{Nannuru2013a}.
A particle filter to approximate the LRFS posterior making use of
a proposal distribution based on the CPHD filter \cite{Nannuru2013b}
for standard superpositional measurements was proposed in \cite{Papi2015}.
\begin{table*}
\centering
\caption{\label{tab:SensorModelProperties}Properties of different TkBD measurement
models}

\begin{tabular}{c|c|c|c|c}
\hline 
Measurement model & \begin{cellvarwidth}[t]
\centering
Measurement density given

the set of targets
\end{cellvarwidth} & Superpositional & \begin{cellvarwidth}[t]
\centering
Gaussian\\
implementation
\end{cellvarwidth} & Examples\tabularnewline
\hline 
General & General & Not required & \XSolidBrush{} & \begin{cellvarwidth}[t]
\centering
\cite{Morelande2007}\cite{F.GarciaFernandez2013}\cite{GarciaFernandez2016}

\cite{Li2018}\cite{Papi2015a}
\end{cellvarwidth}\tabularnewline
\hline 
Standard superpositional & Gaussian & On the conditional mean & \Checkmark{} & \begin{cellvarwidth}[t]
\centering
\cite{Davies2024}\cite{Mahler2014}\cite[eqs. (4), (6)]{Nannuru2013b}

\cite[Sec. III-C]{Papi2015}\cite[eq. (2)]{Nannuru2015}
\end{cellvarwidth}\tabularnewline
\hline 
Extended superpositional & Gaussian & \begin{cellvarwidth}[t]
\centering
On the conditional mean\\
and covariance
\end{cellvarwidth} & \Checkmark{} & \cite{Liang2023}\cite{Zheng2024}\cite{Lynch2025}\tabularnewline
\hline 
General superpositional & General & \begin{cellvarwidth}[t]
\centering
On a parameter of the\\
measurement model
\end{cellvarwidth} & \XSolidBrush{} & \begin{cellvarwidth}[t]
\centering
\cite[eq. (1)]{Nannuru2013b}\cite[Sec. II-B]{Papi2015}

\cite[eq. (1)]{Nannuru2015}
\end{cellvarwidth}\tabularnewline
\hline 
Nested superpositional & \begin{cellvarwidth}[t]
\centering
With tractable conditional \\
mean and covariance
\end{cellvarwidth} & \begin{cellvarwidth}[t]
\centering
On internal variables of \\
the conditional mean\\
and covariance
\end{cellvarwidth} & \Checkmark{} & Proposed in this paper\tabularnewline
\hline 
\end{tabular}
\end{table*}

The Information Exchange Multi-Bernoulli (IEMB) filter \cite{Davies2024}
is a TkBD filter for standard superpositional measurements that propagates
an MB density for the set of targets. The updated MB density minimises
the Kullback-Leibler divergence (KLD) \cite{Bishop2006} between the
exact updated density and its MB approximation, when the target states
are augmented with auxiliary variables. The IEMB filter runs a Bernoulli
filter for each potential target independently, with an exchange of
information among filters to account for the effect of other potential
targets on the mean and covariance matrix of the measurement. This
approach bears similarity to the multiple filtering technique for
a fixed and known number of targets. This technique, used for particle
filtering \cite{Beaudeau2015}\cite{Djuric2007} and Gaussian filtering
\cite{Closas2012}, applies a separate filter to each target accounting
for the mean of the other targets (and possibly the covariance) in
the measurement model.
\begin{figure}
\centering
\includegraphics[width=0.9\columnwidth]{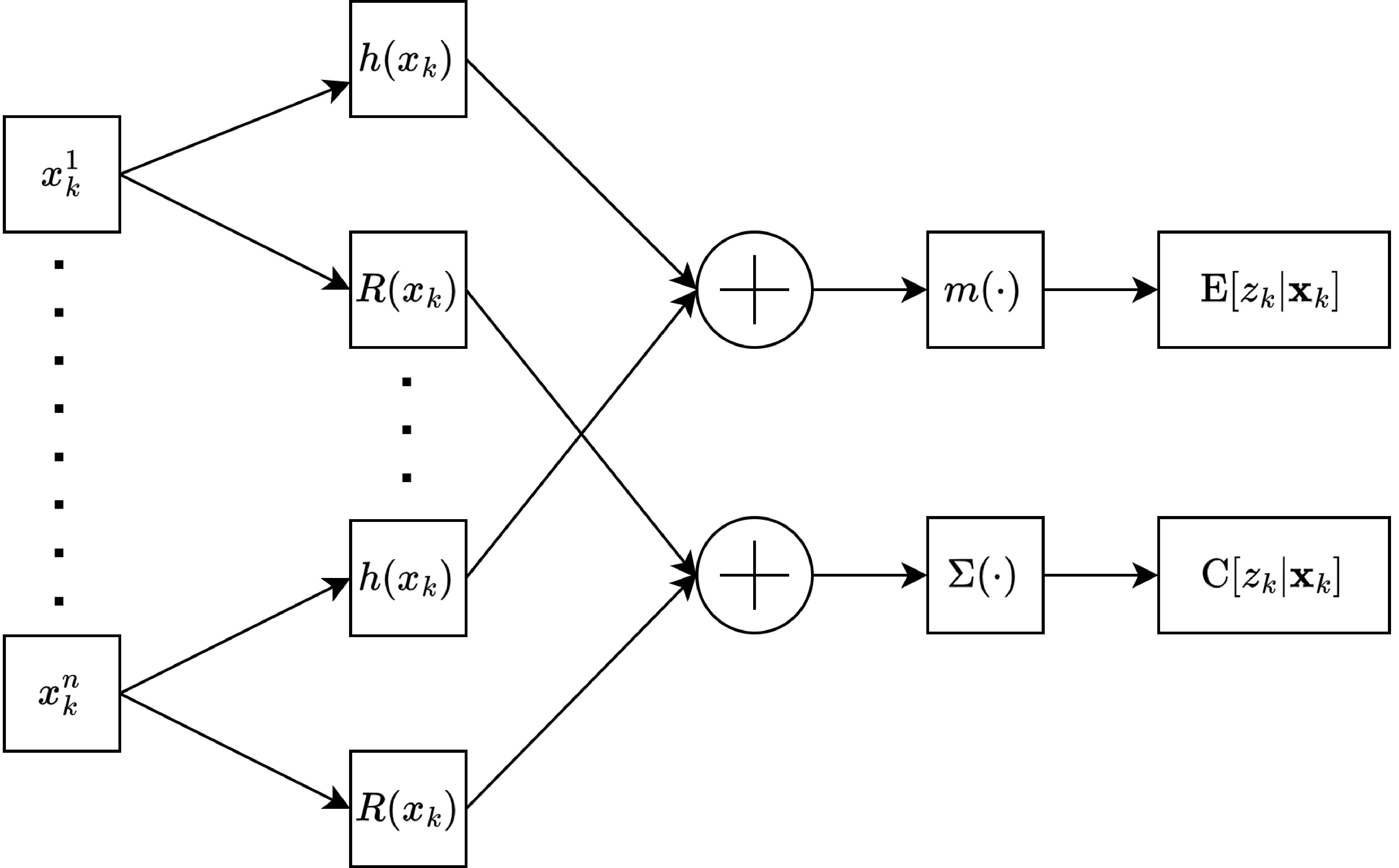}

\caption{\label{fig:MeasurementModelFig}Graphical representation of the nested
superpositional measurement model, with two layers. Given $n$ targets
$\mathbf{x}_{k}=\{x_{k}^{1},...,x_{k}^{n}\}$ at time step $k$, they
contribute to the measurement in the first layer through each hidden
measurement function $h(\cdot)$ and covariance function $R(\cdot)$.
In the second layer these are then summed over all targets and passed
through the mapping functions $m(\cdot)$ and $\Sigma(\cdot)$, providing
the conditional mean and covariance of the measurement.}
\end{figure}

An extended superpositional measurement model that is superpositional
on both the conditional mean and the covariance matrix was proposed
in \cite{Liang2023}. This model can tackle more problems that the
standard superpositional model, and reference \cite{Liang2023} provided
a particle-based belief propagation implementation. Another superpositional
model is what we refer to as a general superpositional measurement
model. In this model, the density of the measurements given the set
of targets depends on a parameter that is formed by the sum (superposition)
of the contributions from each target \cite[eq. (1)]{Nannuru2013b}\cite[Sec. II-B]{Papi2015}\cite[eq. (1)]{Nannuru2015}.
Implementations of TkBD filters based on this model are based on SMC
techniques \cite{Kim2021}.

In many applications, we are interested in estimating target trajectories
or tracks. While one approach to estimate trajectories is to sequentially
link target state estimates with the same label, as the TkBD methods
in \cite{F.GarciaFernandez2013,Li2018,Papi2015a}, a fully Bayesian
approach to estimate target trajectories is based on computing or
approximating the posterior density over the set of trajectories \cite{GarciaFernandez2020a,GarciaFernandez2020c,GarciaFernandez2020,GarciaFernandez2019}
and estimating the set of trajectories directly from the posterior.
\begin{figure}
\centering
\includegraphics[width=1\columnwidth]{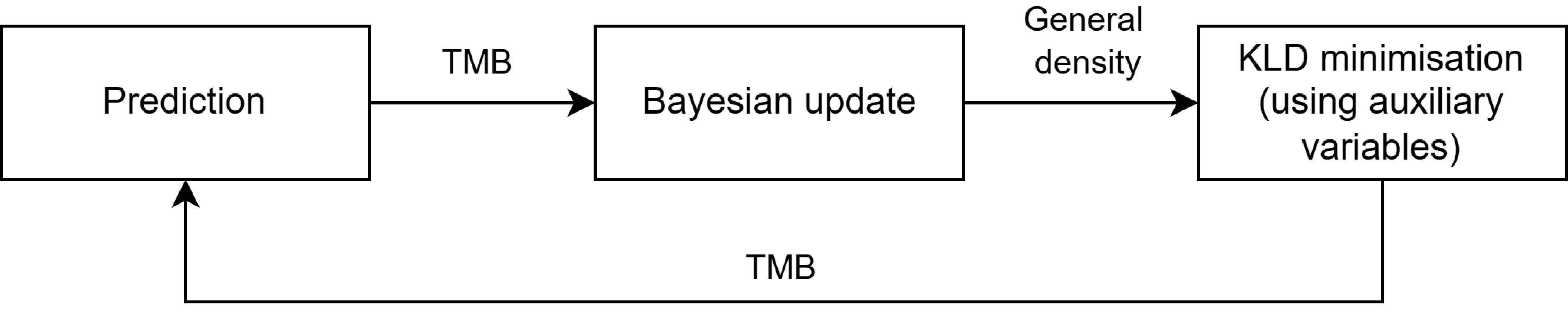}

\caption{\label{fig:T-IEMB-filter-diagram.}T-IEMB filter diagram. A Bayesian
update of the predicted T-MB density and considered likelihood yields
a general non-TMB density. An approximate updated T-MB density is
derived by minimising the KLD (after the introduction of the auxiliary
variables) \cite{Davies2024}.}
\end{figure}

In this paper, we present a TkBD multi-target tracking filter based
on sets of trajectories with nested superpositional measurements.
The first contribution of this paper is the definition of the nested
superpositional measurement model. This model comprises of two layers,
the first containing superpositional hidden variables. These variables
are then mapped, possibly non-linearly, to the second layer which
contains the conditional mean and covariance of the measurement given
the set of targets, as illustrated in Figure \ref{fig:MeasurementModelFig}.
This measurement model is of importance since it generalises the standard
and extended superpositional sensor models, and we can develop computationally
efficient Gaussian filters for TkBD measurement models for which Gaussian
filters have not been derived, see Table \ref{tab:SensorModelProperties}.
It also allows us to model a wide range of measurements with superpositional
properties, which cannot be modelled using Gaussian superpositional
measurements. Such measurement densities include Rician distributed
TkBD measurements in ground radar-based surveillance \cite[Chapter 8]{Mallick2013},
or K-distributed TkBD measurements found in maritime radar-based surveillance
\cite{Ward2013}.

The second contribution of this paper is to develop the IEMB filter
to handle the nested superpositional measurement model (instead of
the standard superpositional model) and to consider sets of trajectories,
giving rise to the Trajectory-IEMB filter (T-IEMB). The T-IEMB filter
propagates a Trajectory-MB (T-MB) density \cite{GarciaFernandez2020c}
for the set of trajectories (either containing the set of alive trajectories,
or the set of all trajectories ever present in the scene). Similarly
to the IEMB filter, the T-IEMB filter update minimises the KLD after
the introduction of auxiliary variables in the single-trajectory state
to obtain a T-MB posterior, as illustrated in Figure \ref{fig:T-IEMB-filter-diagram.}.
Two T-IEMB filters are derived: one with the information on the set
of alive trajectories, and the other with the information on the set
of all trajectories.

The third contribution of the paper is the Gaussian implementation
of the T-IEMB filter for the nested superpositional measurement model.
To do this, we first approximate the conditional moments of the measurement
for each potential target, integrating out the contributions of the
rest of the targets, and using analytical linearisation for the external
mappings in the nested measurement model for a computationally fast
implementation. The Gaussian T-IEMB implementation is then performed
via the iterated posterior linearisation filter (IPLF) for conditional
moments \cite{Tronarp2018}\cite{GarciaFernandez2015}. A side contribution
of the paper is the development of the IEMB filter (for sets of targets)
for the nested superpositional measurement model, obtained by marginalising
out past trajectory states \cite[Thm. 4]{Granstroem2025} in the T-IEMB
filter.

The resulting Gaussian T-IEMB filters can be applied to a wide range
of possibly non-Gaussian TkBD measurements to obtain probabilistic
information on the set of alive and all trajectories. This extends
the capabilities of the IEMB filter \cite{Davies2024} which obtains
information on the set of targets for standard superpositional measurements.
A preliminary version of this work is \cite{Lynch2025}, which presented
the T-IEMB filter for alive trajectories with extended superpositional
measurements.

The remainder of this paper is organised as follows. Section \ref{sec:Models-and-Problem}
details the models used and problem formulation. Section \ref{sec:TrajectoryIEMBFiltering}
presents the T-IEMB filtering recursion for alive and all trajectories.
Section \ref{sec:GaussianIEMBUpdate} presents the update stage of
the Gaussian implementation of the T-IEMB. Section \ref{sec:Simulations}
presents simulation results, and Section \ref{sec:Conclusion} discusses
conclusions of the work.

\section{\label{sec:Models-and-Problem}Models and Problem Formulation}

This paper aims to solve two multi-target tracking problems in TkBD
scenarios with a nested superpositional measurement model: estimating
the set of alive trajectories at the current time step, and estimating
the set of all trajectories up to the current time step \cite{GarciaFernandez2020}.
To achieve this, in this section, we firstly present the multi-target
dynamic model in Section \ref{subsec:Multi-Target-Dynamic-Model},
the nested superpositional measurement model in Section \ref{subsec:GeneralisedMeasurementModel},
and sets of trajectories in Section \ref{subsec:Sets-of-Trajectories}.

\subsection{\label{subsec:Multi-Target-Dynamic-Model}Multi-Target Dynamic Model}

We consider a standard multi-target state and model. An individual
target state $x_{k}\in\mathbb{R}^{n_{x}}$, where $n_{x}$ is the
dimensionality of $x_{k}$, belongs to the set of targets $\mathbf{x}_{k}$
at time step $k$, such that $\mathbf{x}_{k}\in\mathcal{F}(\mathbb{R}^{n_{x}})$,
where $\mathcal{F}(\mathbb{R}^{n_{x}})$ denotes the set of all finite
subsets of $\mathbb{R}^{n_{x}}$. A target with state $x_{k}$ moves
within the surveillance area, surviving to the next time step with
a probability of survival $p^{S}(x_{k})$, with transition density
$g(\cdot|x_{k})$. The multi-target state $\mathbf{x}_{k}$ consists
of both surviving targets and new targets.

New targets are born according to an MB birth model consisting of
$n_{k}^{b}$ birth components, with MB density \cite{GarciaFernandez2020c}
\begin{equation}
f_{k}^{b}(\mathbf{x}_{k})=\sum_{\mathbf{x}^{1}\uplus...\uplus\mathbf{x}^{n_{k}^{b}}=\mathbf{x}_{k}}\prod_{i=1}^{n_{k}^{b}}f_{k}^{b,i}(\mathbf{x}^{i})\label{eq:MBTargetBirthModel}
\end{equation}
where $\uplus$ denotes disjoint union, and the summation sums over
all disjoint and possibly empty subsets $\mathbf{x}^{1},...,\mathbf{x}^{n_{k}^{b}}$
such that their union is $\mathbf{x}_{k}$, which is a result of the
convolution sum for independent RFSs \cite{Mahler2014}. The $i$-th
Bernoulli birth component has a density
\begin{equation}
f_{k}^{b,i}(\mathbf{x}_{k})=\begin{cases}
1-r_{k}^{b,i} & \mathbf{x}_{k}=\emptyset\\
r_{k}^{b,i}p_{k}^{b,i}(x) & \mathbf{x}_{k}=\{x\}\\
0 & \textrm{otherwise}
\end{cases}\label{eq:MBBirthComponents}
\end{equation}
where $\emptyset$ is the empty set, $r_{k}^{b,i}$ is the birth probability,
and $p_{k}^{b,i}(x)$ is the single-target birth density.

\subsection{\label{subsec:GeneralisedMeasurementModel}Nested Superpositional
Measurement Model}

Targets are observed using a sensor system comprising of $M$ sensors,
with the $j$-th sensor generating a measurement $z_{k}^{j}\in\mathbb{R}^{n_{z}}$,
where $j\in\{1,...,M\}$. The measurement from each sensor can be
combined into a measurement vector $z_{k}=[(z_{k}^{1})^{T},...,(z_{k}^{M})^{T}]^{T}$,
where $(\cdot)^{T}$ denotes matrix transpose. In this paper, we use
a nested superpositional measurement model, which we proceed to detail.

The sensor model has a nested architecture, see Figure \ref{fig:MeasurementModelFig}.
Given the set of targets $\mathbf{x}_{k}$ at time step $k$, firstly,
there is an internal (hidden) measurement function $h(\cdot)$, and
internal (hidden) covariance function $R(\cdot)$, such that
\begin{gather}
h(x_{k})=\left[h^{1}(x_{k}),...,h^{M}(x_{k})\right]^{T}\label{eq:Stackedh(x)}\\
R(x_{k})=\textrm{diag}\left(R^{1}(x_{k}),...,R^{M}(x_{k})\right)\label{eq:StackedR(x)}
\end{gather}
where $\textrm{diag}(\cdot)$ denotes a block-diagonal matrix with
the indicated matrices across the diagonal. Then, the conditional
mean and covariance matrix of the measurement vector can be written
as
\begin{gather}
\mathrm{E}[z_{k}|\mathbf{x}_{k}]=m\left(\sum_{x_{k}\in\mathbf{x}_{k}}h(x_{k})\right)\label{eq:GeneralisedMean}\\
\mathrm{C}[z_{k}|\mathbf{x}_{k}]=\Sigma\left(\sum_{x_{k}\in\mathbf{x}_{k}}R(x_{k})\right)\label{eq:GeneralisedCovariance}
\end{gather}
where $m(\cdot)$ is an external mapping which maps the sum of the
internal measurement functions $h(x_{k})$ to a vector of size $Mn_{z}\times1$,
and $\Sigma(\cdot)$ is an external mapping that maps the sum of the
internal covariance functions $R(x_{k})$ to a diagonal covariance
matrix of dimensions $Mn_{z}\times Mn_{z}$.

This nested sensor model provides flexibility in two important properties.
Firstly, the model has a defined set of conditional moments given
the set of targets, which can both be state-dependent. Secondly, we
can model both Gaussian and non-Gaussian measurements, since the external
mappings can be defined to map the hidden functions to the measurement
moments of any arbitrary distribution, with defined mean and covariance
matrix.

We proceed to illustrate the nested measurement model with two examples.
\begin{example}
If the measurement vector $z_{k}\in\mathbb{R}^{M}$ consists of sensor
measurements from targets emitting a signal of fixed energy in a Received
Signal Strength (RSSI) system, the measurements can be modelled as
the multivariate Gaussian \cite{Davies2024}
\begin{align}
p(z_{k}|\mathbf{x}_{k}) & =\mathcal{N}\left(z_{k};\sum_{x_{k}\in\mathbf{x}_{k}}h(x_{k}),R\right)
\end{align}
where $\mathcal{N}(z;\mu,R)$ is a Gaussian density evaluated at $z$,
with mean $\mu$ and covariance matrix $R$, $h(x_{k})$ is the received
power from $x_{k}$ to each sensor, and $R$ is a covariance matrix
which captures the additive noise statistics. The mean and covariance
mapping functions (\ref{eq:GeneralisedMean}) and (\ref{eq:GeneralisedCovariance})
for the measurement vector are
\begin{align}
\mathrm{E}[z_{k}|\mathbf{x}_{k}] & =m\left(\sum_{x_{k}\in\mathbf{x}_{k}}h(x_{k})\right)=\sum_{x_{k}\in\mathbf{x}_{k}}h(x_{k})\\
\mathrm{C}[z_{k}|\mathbf{x}_{k}] & =\Sigma\left(\sum_{x_{k}\in\mathbf{x}_{k}}R(x_{k})\right)=R.
\end{align}
In this case the function $m(\cdot)$ is the identity function, and
the function $\Sigma(\cdot)$ maps to a constant covariance matrix
i.e. it is not state-dependent. Through this selection of the functions
$m(\cdot)$ and $\Sigma(\cdot)$, we have recovered the standard superpositional
measurement model in \cite{Davies2024}.
\end{example}
We now present a non-Gaussian example, which demonstrates the flexibility
of the proposed measurement model.
\begin{example}
We consider a single sensor $M=1$ for simplicity of exposition. The
measurement $z_{k}\in\mathbb{R}$ originates from a Swerling 1 target
in a radar system, whose Radar Cross Section (RCS) is modelled as
a random variable. The density of the measurement is Rayleigh distributed
as \cite{Morelande2007}
\begin{equation}
p(z_{k}|\mathbf{x}_{k})=\frac{z_{k}}{1+\lambda(\mathbf{x}_{k})}\exp\left(-\frac{(z_{k})^{2}}{2(1+\lambda(\mathbf{x}_{k}))}\right)\label{eq:RayleighAmplitudeMeasurement}
\end{equation}
where $\lambda(\mathbf{x}_{k})=\sum_{x_{k}\in\mathbf{x}_{k}}h(x_{k})$
gives the Signal-to-Noise Ratio (SNR) accounting for all targets.
The mean and covariance mapping functions (\ref{eq:GeneralisedMean})
and (\ref{eq:GeneralisedCovariance}) for the measurement are \cite{Papoulis2002}
\begin{align}
\mathrm{E}[z_{k}|\mathbf{x}_{k}] & =m\left(\sum_{x_{k}\in\mathbf{x}_{k}}h(x_{k})\right)=\sqrt{\frac{\pi}{2}(1+\lambda(\mathbf{x}_{k}))}\label{eq:ExampleRayleighMean}\\
\mathrm{C}[z_{k}|\mathbf{x}_{k}] & =\Sigma\left(\sum_{x_{k}\in\mathbf{x}_{k}}R(x_{k})\right)=\frac{4-\pi}{2}(1+\lambda(\mathbf{x}_{k}))\label{eq:ExampleRayleighCovariance}
\end{align}
where $\sum_{x_{k}\in\mathbf{x}_{k}}R(x_{k})=\lambda(\mathbf{x}_{k})$.

Another example of the nested measurement model with the Rice distribution
is considered in the simulations in Section \ref{sec:Simulations}.
\end{example}

\subsection{\label{subsec:Sets-of-Trajectories}Sets of Trajectories}

In this paper we aim to infer sets of trajectories, arising from the
multi-target model in Section \ref{subsec:Multi-Target-Dynamic-Model},
and from measurements distributed according to the nested model in
Section \ref{subsec:GeneralisedMeasurementModel}. We consider the
single-trajectory variable $X=(t,x^{1:v})$, where $t$ is the time
step at which the trajectory began, and $x^{1:v}=(x^{1},...,x^{v})$
is the sequence of target states of length $v$. The start time and
length $(t,v)$ therefore belong to the set $I_{(k)}=\{(t,v):0\leq t\leq k\ \textrm{and}\ 1\leq v\leq k-t+1\}$,
and the single-trajectory space up to the current time step $k$ is
then $T_{(k)}=\uplus_{(t,v)\in I_{(k)}}\{t\}\times\mathbb{R}^{vn_{x}}$
\cite{GarciaFernandez2020a}. A set of trajectories up to the current
time step $k$ is denoted by $\mathbf{X}_{k}$, where the set of trajectories
$\mathbf{X}_{k}\in\mathcal{F}(T_{(k)})$.

We use $\tau^{k}(X)$ to denote the single-target at time step $k$
corresponding to trajectory $X$. This set may be empty or contain
one element with the expression
\begin{equation}
\tau^{k}(t,x^{1:v})=\begin{cases}
\{x^{k-t+1}\} & t\leq k\leq t+v-1\\
\emptyset & \textrm{elsewhere}.
\end{cases}
\end{equation}
Similarly, $\mathbf{x}_{k}$ denotes the set of targets at time step
$k$ corresponding to the set of trajectories $\mathbf{X}_{k}$. Integration
over sets of trajectories is reviewed in Appendix A.

In a Bayesian framework, all information about the set of considered
trajectories (either the alive or all trajectories \cite{GarciaFernandez2020})
is included in its posterior density, which can be written in terms
of the prediction and update steps \cite{GarciaFernandez2020a}. Since
the update step is intractable for the measurement model in Section
\ref{subsec:GeneralisedMeasurementModel}, we propose an approximate
T-MB filter in the next section.

\section{\label{sec:TrajectoryIEMBFiltering}Trajectory IEMB Filter}

In this section we begin by detailing the dynamic model for the set
of alive and all trajectories in Section \ref{subsec:TrajDynamicModel},
before presenting the T-MB density in Section \ref{subsec:Trajectory-Multi-Bernoulli-Densi},
which is propagated through the T-IEMB prediction and update recursions
in Section \ref{subsec:T-IEMB-Prediction} and \ref{subsec:T-IEMB-Update},
respectively.

\subsection{\label{subsec:TrajDynamicModel}Dynamic Models for Sets of Trajectories}

Sets of alive and all trajectories evolve according to the following
dynamic models, which are derived from the dynamic model for the sets
of targets \cite{GarciaFernandez2020c}. Following (\ref{eq:MBTargetBirthModel}),
new trajectories are born at time step $k$ according to an MB birth
model for trajectories
\begin{equation}
f_{k}^{b}(\mathbf{X}_{k})=\sum_{\mathbf{X}^{1}\uplus...\uplus\mathbf{X}^{n_{k}^{b}}=\mathbf{X}_{k}}\prod_{i=1}^{n_{k}^{b}}f_{k}^{b,i}(\mathbf{X}^{i})\label{eq:MBTrajBirth}
\end{equation}
where each Bernoulli component is now a trajectory of length 1 born
at time step $k$
\begin{equation}
f_{k}^{b,i}(\mathbf{X}_{k})=\begin{cases}
1-r_{k}^{b,i} & \mathbf{X}_{k}=\emptyset\\
r_{k}^{b,i}p_{k}^{b,i}(x) & \mathbf{X}_{k}=\{(k,x)\}\\
0 & \textrm{otherwise}.
\end{cases}\label{eq:MBTrajBirthDensity}
\end{equation}
Note that this density corresponds to the birth model for sets of
targets, adding the time index $k$ to the state. A trajectory $X$
survives to the next time step with probability $p^{S}(X)$ and evolves
with a single-trajectory density $g_{k+1}(\cdot|X)$. The single-trajectory
dynamic model differs if we consider alive or all trajectories.

\subsubsection{Dynamic Model for Sets of Alive Trajectories}

When estimating the set of alive trajectories, the filtering density
at any time step contains information on only the trajectories that
are alive. At any time step $k$, each alive trajectory $X=(t,x^{1:v})$
in the set of alive trajectories $\mathbf{X}_{k}$, survives to the
next time step with probability $p^{S}(X)=p^{S}(x^{v})$, and evolves
according to the single-trajectory transition density \cite{GarciaFernandez2020}
\begin{multline}
g_{k+1}(t_{y},y^{1:v_{y}}|X)=\delta_{t}[t_{y}]\delta_{v+1}[v_{y}]\delta_{x^{1:v}}(y^{1:v_{y}-1})\\
\times g(y^{v_{y}}|x^{v})\label{eq:AliveTrajTransition}
\end{multline}
where the Kronecker delta $\delta_{t}[t_{y}]=1$ if $t=t_{y}$, and
$\delta_{x^{1:v}}(y^{1:v_{y}-1})$ is the Dirac delta located at $x^{1:v}$
and evaluated at $y^{1:v_{y}-1}$. Trajectories alternatively die
with probability $1-p^{S}(x^{v})$.

The set of alive trajectories at time step $k+1$ is then the union
of surviving trajectories, which have evolved according to (\ref{eq:AliveTrajTransition}),
and new trajectories which have been born according to the MB birth
model (\ref{eq:MBTrajBirthDensity}).

\subsubsection{Dynamic Model for Sets of All Trajectories}

The filtering density for sets of all trajectories contains both alive
trajectories, and trajectories which have died at a previous time
step and are no longer evolving. Therefore, at any time step $k$,
each trajectory $X=(t,x^{1:v})$ in the set of all trajectories $\mathbf{X}_{k}$
survives to the next time step with probability $p^{S}(X)=1$, with
single-trajectory transition density \cite{GarciaFernandez2020}
\begin{multline}
g_{k+1}(t_{y},y^{1:v_{y}}|X)=\delta_{t}[t_{y}]\\
\times\begin{cases}
\delta_{v}[v_{y}]\delta_{x^{1:v}}(y^{1:v_{y}}) & \omega_{y}<k\\
\delta_{v}[v_{y}]\delta_{x^{1:v}}(y^{1:v_{y}})(1-p^{S}(x^{v})) & \omega_{y}=k\\
\delta_{v+1}[v_{y}]\delta_{x^{1:v}}(y^{1:v_{y}-1})p^{S}(x^{v})\\
\times g(y^{v_{y}}|x^{v}) & \omega_{y}=k+1
\end{cases}\label{eq:AllTrajTransition}
\end{multline}
where $\omega_{y}=t_{y}+v_{y}-1$. That is, each trajectory stays
in the set of trajectories with probability 1, and its length increases
with probability $p^{S}(x^{v})$. New trajectories are born according
to the same MB birth model (\ref{eq:MBTrajBirthDensity}) as for the
set of alive trajectories.

\subsection{\label{subsec:Trajectory-Multi-Bernoulli-Densi}Trajectory Multi-Bernoulli
Density}

The proposed TkBD filter propagates MB densities on the set of trajectories
through the filtering recursion. That is, the predicted and posterior
densities at time step $k$ given the measurements up to time step
$k'\in\{k-1,k\}$ are T-MB of the form \cite{GarciaFernandez2020c}
\begin{equation}
f_{k|k'}(\mathbf{X}_{k})=\sum_{\biguplus_{i=1}^{n_{k|k'}}\mathbf{X}^{i}=\mathbf{X}_{k}}\prod_{i=1}^{n_{k|k'}}f_{k|k'}^{i}(\mathbf{X}^{i})\label{eq:TrajMBDensity}
\end{equation}
where $n_{k|k'}$ is the number of Bernoulli components, and
\begin{equation}
f_{k|k'}^{i}(\mathbf{X}_{k})=\begin{cases}
1-r_{k|k'}^{i} & \mathbf{X}_{k}=\emptyset\\
r_{k|k'}^{i}p_{k|k'}^{i}(X) & \mathbf{X}_{k}=\{X\}\\
0 & \textrm{otherwise}
\end{cases}\label{eq:TrajBernoulliDensity}
\end{equation}
where $r_{k|k'}^{i}$ is the probability of existence of the $i$-th
Bernoulli component, and $p_{k|k'}^{i}(X)$ is the single-trajectory
density, which has a deterministic known start time $t^{i}$.

\subsection{\label{subsec:T-IEMB-Prediction}T-IEMB Filter Prediction}

The T-IEMB filter prediction is equivalent to the T-MB filter prediction
\cite{GarciaFernandez2020c}, since the dynamic models are the same.
\begin{lem}[T-IEMB Prediction]
 Given a T-MB posterior of the form (\ref{eq:TrajMBDensity}), the
predicted density is T-MB of the form (\ref{eq:TrajMBDensity}) with
$n_{k|k-1}=n_{k-1|k-1}+n_{k}^{b}$ components, where the probability
of existence and single-trajectory density of each $i\in\{1,...,n_{k-1|k-1}\}$
Bernoulli is \cite{GarciaFernandez2020c}
\begin{gather}
r_{k|k-1}^{i}=r_{k-1|k-1}^{i}\int p^{S}(X')p_{k-1|k-1}^{i}(X')dX'\\
p_{k|k-1}^{i}(X)=\frac{\int g_{k}(X|X')p^{S}(X')p_{k-1|k-1}^{i}(X')dX'}{\int p^{S}(X')p_{k-1|k-1}^{i}(X')dX'}
\end{gather}
where the integral is a single-trajectory integral (see Appendix A),
$g_{k}(X|X')$ is the transition density (\ref{eq:AliveTrajTransition})
for alive trajectories, and (\ref{eq:AllTrajTransition}) for all
trajectories. For each $i\in\{n_{k-1|k-1}+1,...,n_{k-1|k-1}+n_{k}^{b}\}$,
its Bernoulli birth component is
\begin{equation}
f_{k|k-1}^{i}(\mathbf{X})=f_{k}^{b,i-n_{k-1|k-1}}(\mathbf{X}).
\end{equation}
\end{lem}

\subsection{\label{subsec:T-IEMB-Update}T-IEMB Filter Update}

To present the T-IEMB filter update, we first present a more explicit
form of the single-trajectory density.

\subsubsection{Form of the Single-Trajectory Density}

The predicted and updated single-trajectory densities in the T-IEMB
filter are of the form
\begin{equation}
p_{k|k'}^{i}(X)=\sum_{l=t^{i}}^{k}\beta_{k|k'}^{i}(l)p_{k|k'}^{i,l}(X)\label{eq:GeneralTrajForm}
\end{equation}
where $\beta_{k|k'}^{i}(l)$ corresponds to the probability that the
trajectory ends at time step $l$, and $p_{k|k'}^{i,l}(X)$ is a single-trajectory
density ending at time step $l$. For the set of alive trajectories,
$\beta_{k|k'}^{i}(k)=1$ and $\beta_{k|k'}^{i}(l)=0$ for $l\in\{t^{i},...,k-1\}$,
as we only consider trajectories that are alive at time step $k$.
For the set of all trajectories, we consider trajectories that may
end at any time step $l\in\{t^{i},...,k\}$, meaning each $p_{k|k'}^{i,l}(X)$
is weighted by its corresponding $\beta_{k|k'}^{i}(l)$. This results
in a weighted mixture density where $\sum_{l=t^{i}}^{k}\beta_{k|k'}^{i}(l)=1$.

\subsubsection{Auxiliary Variables}

To obtain the T-IEMB update, we follow the IEMB filter update \cite{Davies2024}
which makes use of auxiliary variables, which are a common probabilistic
inference tool, see auxiliary particle filtering \cite{Pitt1999},
expectation maximisation \cite{Bishop2006} and factor graph opening
\cite{Wymeersch_2007}. In the MTT context, auxiliary variables represent
track indices that are implicit in the posterior (but they are not
part of the birth, dynamic and measurement models). These have been
used in other RFS-based MTT \cite{GarciaFernandez2020c}\cite{GarciaFernandez2020}\cite{GarciaFernandez2026}
to remove the convolution sum in the posterior, see (\ref{eq:TrajMBDensity}),
and enable closed-form KLD minimisations.

Auxiliary variables are not necessarily unique, since for a Poisson
point process birth model, there can be multiple targets with the
same auxiliary variable representing undetected targets \cite{GarciaFernandez2020}.
This property makes them different from target labels used in labelled
RFSs \cite{Vo2013}. Nevertheless, for the case of MB birth, this
definition of auxiliary variables can be considered equivalent to
labels, since they are unique and each is representing a Bernoulli
component at the time of birth, though they are not part of the birth,
dynamic and measurement models.

We introduce an auxiliary variable $u$ to the single-trajectory space,
such that the single-trajectory state is $(u,X)$. Each $u$ takes
a value in the space $\mathbb{U}_{k}=\{1,...,n_{k|k-1}\}$, and the
single-trajectory state space then becomes $\mathbb{U}_{k}\times T_{(k)}$.
A set of trajectories with auxiliary variables is denoted by $\mathbf{\widetilde{X}}_{k}$,
and we define that the set of trajectories with auxiliary variable
$i$ (with at most one trajectory) is
\begin{equation}
\mathbf{\widetilde{X}}_{k}^{i}=\left\{ (u,X)\in\mathbf{\widetilde{X}}_{k}:u=i\right\} .
\end{equation}
The predicted T-MB density with auxiliary variables is then
\begin{equation}
\widetilde{f}_{k|k-1}(\mathbf{\widetilde{X}}_{k})=\prod_{i=1}^{n_{k|k-1}}\widetilde{f}_{k|k-1}^{i}(\mathbf{\widetilde{X}}_{k}^{i})\label{eq:PredictedMBAux}
\end{equation}
where each Bernoulli component with auxiliary variables is \cite{GarciaFernandez2020}
\begin{equation}
\widetilde{f}_{k|k-1}^{i}(\mathbf{\widetilde{X}}_{k})=\begin{cases}
1-r_{k|k-1}^{i} & \mathbf{\mathbf{\widetilde{X}}}_{k}=\emptyset\\
r_{k|k-1}^{i}p_{k|k-1}^{i}(X)\delta_{i}[u] & \mathbf{\mathbf{\widetilde{X}}}_{k}=\{(u,X)\}\\
0 & \textrm{otherwise}
\end{cases}\label{eq:AuxiliaryBernoulli}
\end{equation}
and the predicted T-MB density of the form (\ref{eq:TrajMBDensity})
reduces to (\ref{eq:PredictedMBAux}) since the auxiliary variables
ensure there is only one disjoint union of subsets of $\mathbf{\widetilde{X}}_{k}$.
We then aim to obtain an approximate T-MB posterior of the form
\begin{equation}
\widetilde{f}_{k|k}(\mathbf{\widetilde{X}}_{k})=\prod_{i=1}^{n_{k|k}}\widetilde{f}_{k|k}^{i}(\mathbf{\widetilde{X}}_{k}^{i})\label{eq:MBApprox}
\end{equation}
where each Bernoulli component $\widetilde{f}_{k|k}^{i}(\mathbf{\widetilde{X}}_{k}^{i})$
is of the form (\ref{eq:AuxiliaryBernoulli}).

\subsubsection{Update}

Given a predicted T-MB density of the form (\ref{eq:TrajMBDensity}),
the posterior density can be found by applying a Bayesian update consisting
of the prior and likelihood densities, which we write as
\begin{equation}
f_{k|k}(\mathbf{X}_{k})=\frac{p(z_{k}|\mathbf{x}_{k})f_{k|k-1}(\mathbf{X}_{k})}{\int\left[p(z_{k}|\mathbf{x}_{k})f_{k|k-1}(\mathbf{X}_{k})\right]\delta\mathbf{X}_{k}}\label{eq:TruePosterior}
\end{equation}
where the integral is the set integral for trajectories (see Appendix
A) \cite{GarciaFernandez2020a}. The density (\ref{eq:TruePosterior})
is not T-MB, therefore the T-IEMB update stage computes a T-MB approximation
to the true non-TMB density (\ref{eq:TruePosterior}).

Similarly to the proof for sets of targets \cite[Appendix A]{Davies2024},
the MB approximation (\ref{eq:MBApprox}) which minimises the KLD
between it and (\ref{eq:TruePosterior}) (after adding the auxiliary
variables), has Bernoulli components given by
\begin{equation}
\widetilde{f}_{k|k}^{u}(\mathbf{\widetilde{X}}_{k}^{u})=\frac{p^{u}(z_{k}|\mathbf{\widetilde{x}}_{k}^{u})\tilde{f}_{k|k-1}(\mathbf{\widetilde{X}}_{k}^{u})}{\int p^{u}(z_{k}|\mathbf{\widetilde{x}}_{k}^{u})\tilde{f}_{k|k-1}(\mathbf{\widetilde{X}}_{k}^{u})\delta\mathbf{\widetilde{X}}_{k}^{u}}\label{eq:MBApproxComponent}
\end{equation}
where $p^{u}(z_{k}|\mathbf{\widetilde{x}}_{k}^{u})$ is a modified
likelihood for the $u$-th potential target, given by
\begin{equation}
p^{u}(z_{k}|\mathbf{\widetilde{x}}_{k}^{u})=\int p(z_{k}|\mathbf{\widetilde{x}}_{k})\prod_{i=1:i\neq u}^{n_{k|k-1}}\widetilde{f}_{k|k-1}^{i}(\mathbf{\widetilde{x}}_{k}^{i})\delta\mathbf{\widetilde{x}}_{k}^{(-u)}\label{eq:IntegratedLikelihood}
\end{equation}
where $\widetilde{\mathbf{x}}_{k}=\widetilde{\mathbf{x}}_{k}^{1}\cup...\cup\mathbf{\widetilde{x}}_{k}^{n_{k|k-1}}$
is the set of all targets and $\mathbf{\widetilde{x}}_{k}^{(-u)}=(\mathbf{\widetilde{x}}_{k}^{1},...,\mathbf{\widetilde{x}}_{k}^{u-1},\mathbf{\widetilde{x}}_{k}^{u+1},...\mathbf{\widetilde{x}}_{k}^{n_{k|k-1}})$
is the sequence of sets of all targets excluding the $u$-th target.
(\ref{eq:IntegratedLikelihood}) integrates out the information from
all Bernoulli components except the $u$-th one.

Each predicted single-target Bernoulli $\widetilde{f}_{k|k-1}^{i}(\mathbf{\widetilde{x}}_{k}^{i})$
in (\ref{eq:IntegratedLikelihood}) can be found from the corresponding
trajectory Bernoulli $\widetilde{f}_{k|k-1}^{i}(\mathbf{\widetilde{X}}_{k}^{i})$
with single-trajectory density (\ref{eq:GeneralTrajForm}) using the
marginalisation theorem for T-MB densities \cite[Theorem 4]{Granstroem2025},
which \foreignlanguage{british}{gives
\begin{align}
\widetilde{f}_{k|k-1}^{i}(\widetilde{\mathbf{x}}_{k}^{i}) & =\begin{cases}
r_{k|k-1}^{i}(k)p_{k|k-1}^{i}(x) & \widetilde{\mathbf{x}}_{k}^{i}=\{(i,x)\}\\
1-r_{k|k-1}^{i}(k) & \widetilde{\mathbf{x}}_{k}^{i}=\emptyset\\
0 & \mathrm{otherwise}
\end{cases}
\end{align}
where the probability of existence of the target at time step $k$
is
\begin{align}
r_{k|k-1}^{i}(k) & =r_{k|k-1}^{i}\beta_{k|k-1}^{i}(k)\label{eq:CurrentTimeExistenceProb}
\end{align}
and the single-target density is
\begin{align}
p_{k|k-1}^{i}(x) & =\int p_{k|k-1}^{i,k}(t^{i},y^{1:k-t^{i}},x)dy^{1:k-t^{i}}
\end{align}
}which integrates out all previous target states in the single-trajectory
density component that ends at time step $k$.

\noindent We now present the update for alive trajectories.
\begin{lem}[T-IEMB Update for Alive Trajectories]
\label{lem:-Given-a-1} Given a predicted T-MB density of the form
(\ref{eq:PredictedMBAux}), the updated density is T-MB of the form
(\ref{eq:MBApprox}), where the updated single-trajectory density
and probability of existence of the $u$-th Bernoulli is given by
\begin{gather}
p_{k|k}^{u}(X)=\frac{p^{u}(z_{k}|\tau^{k}(X))p_{k|k-1}^{u,k}(X)}{\int p^{u}(z_{k}|\tau^{k}(X))p_{k|k-1}^{u,k}(X)dX}\\
r_{k|k}^{u}=\frac{p^{u}(z_{k}|\lvert\widetilde{\mathbf{x}}_{k}^{u}|=1)r_{k|k-1}^{u}}{p^{u}(z_{k}|\lvert\widetilde{\mathbf{x}}_{k}^{u}|=1)r_{k|k-1}^{u}+p^{u}\left(z_{k}|\emptyset\right)(1-r_{k|k-1}^{u})}\label{eq:ProbabilityExistenceAlive}
\end{gather}
where $|\cdot|$ denotes the cardinality of a set, and $p^{u}(z_{k}|\lvert\widetilde{\mathbf{x}}_{k}^{u}|=1)$
is the likelihood of target existence
\begin{equation}
p^{u}(z_{k}|\lvert\widetilde{\mathbf{x}}_{k}^{u}|=1)=\int p^{u}(z_{k}|\{x\})p_{k|k-1}^{u}(x)dx\label{eq:ExistenceLikelihood}
\end{equation}
and $p^{u}(z_{k}|\emptyset)$ is given by (\ref{eq:IntegratedLikelihood})
with $\mathbf{\widetilde{x}}_{k}^{u}=\emptyset$.
\end{lem}
\noindent We now present the update for all trajectories.
\begin{lem}[T-IEMB Update for All Trajectories]
\label{lem:-Given-a} Given a predicted T-MB density of the form
(\ref{eq:PredictedMBAux}), the updated density is T-MB of the form
(\ref{eq:MBApprox}), where the updated single-trajectory density,
updated probability of existence, and updated probability of the trajectory
ending at time step $l\in\{t^{u},...,k\}$ for the $u$-th Bernoulli
is given by
\begin{gather}
p_{k|k}^{u,l}(X)=\begin{cases}
p_{k|k-1}^{u,l}(X) & l\in\{t^{u},...,k-1\}\\
\frac{p^{u}(z_{k}|\tau^{k}(X))p_{k|k-1}^{u,l}(X)}{\int p^{u}(z_{k}|\tau^{k}(X))p_{k|k-1}^{u,l}(X)dX} & l=k
\end{cases}\\
r_{k|k}^{u}=\frac{\rho_{k}^{u}r_{k|k-1}^{u}}{\rho_{k}^{u}r_{k|k-1}^{u}+(1-r_{k|k-1}^{u})p^{u}\left(z_{k}|\emptyset\right)}\label{eq:ProbabilityExistenceAll}\\
\beta_{k|k}^{u}(l)\propto\begin{cases}
p^{u}(z_{k}|\emptyset)\beta_{k|k-1}^{u}(l) & l\in\{t^{u},...,k-1\}\\
p^{u}(z_{k}||\mathbf{\widetilde{x}}_{k}^{i}|=1)\beta_{k|k-1}^{u}(l) & l=k
\end{cases}\label{eq:UpdateBeta}
\end{gather}
where
\begin{equation}
\rho_{k}^{u}=p^{u}(z_{k}||\mathbf{\widetilde{x}}_{k}^{u}|=1)\beta_{k|k-1}^{u}(k)+p^{u}(z_{k}|\emptyset)\sum_{l=t^{u}}^{k-1}\beta_{k|k-1}^{u}(l)
\end{equation}
and $p^{u}(z_{k}|\lvert\widetilde{\mathbf{x}}_{k}^{u}|=1)$ is given
by (\ref{eq:ExistenceLikelihood}) and $p^{u}(z_{k}|\emptyset)$ by
(\ref{eq:IntegratedLikelihood}) with $\mathbf{\widetilde{x}}_{k}^{u}=\emptyset$.
\end{lem}
Lemma \ref{lem:-Given-a} is proved in Appendix B. It should be noted
that if $\beta_{k|k-1}^{u}(k)=1$, which corresponds to the case of
alive trajectories and implies that $\beta_{k|k}^{u}(l)=0$ for $l\in\{t^{u},...,k-1\}$,
then Lemma \ref{lem:-Given-a} reduces to Lemma \ref{lem:-Given-a-1},
which is also proved.

We also note that $\beta_{k|k}^{u}(k)$ is proportional to its predicted
value $\beta_{k|k-1}^{u}(k)$ multiplied by the likelihood of the
potential target being present, since $\beta_{k|k}^{u}(k)$ represents
the situation in which the trajectory has existed and is currently
alive. On the contrary, $\beta_{k|k}^{u}(l)$ for $l\in\{t^{u},...,k-1\}$,
is proportional to its predicted value times the likelihood of the
potential target not existing, since in this case, $\beta_{k|k}^{u}(l)$
represents the situation in which the trajectory has existed but ended
before the current time step, so it contributes to the likelihood
with the empty set.

\section{\label{sec:GaussianIEMBUpdate}Gaussian T-IEMB Filter}

We proceed to present the Gaussian implementation of the T-IEMB, termed
GT-IEMB, for estimating the set of alive and all trajectories with
the nested superpositional measurement model of Section \ref{subsec:GeneralisedMeasurementModel}.
In this section we focus on the update as this is where our contributions
lie. The prediction stage corresponds to the Gaussian implementation
of the T-MB filter \cite{GarciaFernandez2020c} \cite{GarciaFernandez2020},
which we include for alive and all trajectories in Appendix C for
completeness.

We firstly detail the Gaussian trajectory densities in Section \ref{subsec:GaussianModels},
along with the conditional moments of the nested superpositional measurement
model in Section \ref{subsec:Conditional-Moments-of}. The conditional
moments are used to develop the single-target Gaussian update in Section
\ref{subsec:GaussianUpdate}, which returns the updated single-target
moments. Section \ref{subsec:Single-Trajectory-Gaussian} explains
how single-trajectories can be updated using these. Finally, Sections
\ref{subsec:Practical-Considerations} and \ref{subsec:Estimation}
discuss practical considerations and trajectory estimation, respectively.

\subsection{\label{subsec:GaussianModels}Gaussian Trajectory Densities}

A Gaussian trajectory density with start time $\tau$ and duration
$\iota$ is denoted by \cite{GarciaFernandez2020}
\begin{equation}
\mathcal{N}(t,x^{1:v};\tau,\overline{x},P)=\begin{cases}
\mathcal{N}(x^{1:v};\overline{x},P) & t=\tau,v=\iota\\
0 & \textrm{otherwise}
\end{cases}
\end{equation}
where the duration $\iota=\textrm{dim}(\overline{x})/n_{x}$, where
$\textrm{dim}(\cdot)$ denotes dimensionality, and the Gaussian trajectory
density is evaluated at $(t^{i},x^{1:v})$, with mean $\overline{x}\in\mathbb{R}^{\iota n_{x}}$
and covariance $P\in\mathbb{R}^{\iota n_{x}}\times\mathbb{R}^{\iota n_{x}}$.

\subsubsection{Gaussian Single-Trajectory Density}

The Gaussian single-trajectory density (\ref{eq:GeneralTrajForm})
of the $i$-th Bernoulli component in (\ref{eq:TrajMBDensity}) is
of the form
\begin{equation}
p_{k|k'}^{i}(X)=\sum_{l=t^{i}}^{k}\beta_{k|k'}^{i}(l)\mathcal{N}(X;t^{i},\overline{x}_{k|k'}^{i}(l),P_{k|k'}^{i}(l))\label{eq:GaussianSingleTraj}
\end{equation}
where $t^{i}$ is the trajectory start time, $\overline{x}_{k|k'}^{i}(l)\in\mathbb{R}^{\iota n_{x}}$
is the trajectory mean ending at time step $l$, and $P_{k|k'}^{i}(l)\in\mathbb{R}^{\iota n_{x}}\times\mathbb{R}^{\iota n_{x}}$
is the covariance matrix for the trajectory ending at time step $l$,
where $\iota=l-t^{i}+1$. For the set of alive trajectories, $\beta_{k|k'}^{i}(k)=1$
and $\beta_{k|k'}^{i}(l)=0$ for $l\in\{t^{i},...,k-1\}$. For the
set of all trajectories, we have a weighted Gaussian mixture density
where $\sum_{l=t^{i}}^{k}\beta_{k|k'}^{i}(l)=1$.

\subsubsection{Gaussian Single-Target Density}

\selectlanguage{british}%
The T-IEMB filter updates for alive and all trajectories (Lemmas \ref{lem:-Given-a-1}
and \ref{lem:-Given-a}) require updating the single-trajectory density
component that is alive, denoted by $p_{k|k-1}^{i,k}(X)$, while the
rest of the single-trajectory density components remain unchanged.
In the Gaussian implementation, $p_{k|k-1}^{i,k}(X)$ has trajectory
mean $\overline{x}_{k|k-1}^{i}(k)$ and covariance matrix $P_{k|k-1}^{i}(k)$.

As the measurement likelihood depends only on the current single-target
states, see Section (\ref{subsec:GeneralisedMeasurementModel}), we
perform the update by first updating the single-target density $p_{k|k-1}^{i,k}(x)$
at time step $k$ associated with the single-trajectory density $p_{k|k-1}^{i,k}(X)$.
This density is Gaussian with a density
\begin{align}
p_{k|k-1}^{i,k}(x) & =\mathcal{N}(x;\mu_{k|k-1}^{i},\Xi_{k|k-1}^{i})
\end{align}
where its mean $\mu_{k|k-1}^{i}\in\mathbb{R}^{n_{x}}$ and covariance
matrix $\Xi_{k|k-1}^{i}\in\mathbb{R}^{n_{x}\times n_{x}}$ are directly
obtained from the trajectory mean and covariance matrix. We explain
how to compute the updated mean $\mu_{k|k}^{i}$ and covariance $\Xi_{k|k}^{i}$
in Section \ref{subsec:GaussianUpdate}, for which we first require
the conditional moments of the measurement in Section \ref{subsec:Conditional-Moments-of}.
Finally, we explain how to update past states of the trajectories
in Section \ref{subsec:Single-Trajectory-Gaussian}.
\selectlanguage{english}%

\subsection{\label{subsec:Conditional-Moments-of}Conditional Moments of the
Measurement}

The Gaussian update for the $u$-th target, following \cite{Tronarp2018},
requires the conditional measurement mean and covariance given the
target set $\widetilde{\mathbf{x}}^{u}$. We firstly introduce the
following variables and their moments that will be required for the
update. Given a single-target $\widetilde{\mathbf{x}}^{i}$, we define
these internal variables related to the nested measurement model
\begin{gather}
h^{i}=\begin{cases}
h(x) & \widetilde{\mathbf{x}}^{i}=\{(i,x)\}\\
0 & \widetilde{\mathbf{x}}^{i}=\emptyset
\end{cases}\\
R^{i}=\begin{cases}
R(x) & \widetilde{\mathbf{x}}^{i}=\{(i,x)\}\\
0 & \widetilde{\mathbf{x}}^{i}=\emptyset.
\end{cases}
\end{gather}

\begin{lem}
\label{lem:Moments_of_y^i}Given an MB predicted density of the form
(\ref{eq:TrajMBDensity}) with single-trajectory densities of the
form (\ref{eq:GaussianSingleTraj}), the mean and covariance of $h^{i}$,
and the mean of $R^{i}$, are
\begin{align}
\mathrm{E}[h^{i}] & =r_{k|k-1}^{i}(k)\mathrm{E}^{i}[h(x)]\label{eq:Meany^i}\\
\mathrm{C}[h^{i}] & =r_{k|k-1}^{i}(k)\mathrm{E}^{i}[h(x)h(x)^{T}]\nonumber \\
 & -(r_{k|k-1}^{i}(k))^{2}\mathrm{E}^{i}[h(x)](\mathrm{E}^{i}[h(x)])^{T}\label{eq:Covariancey^i}\\
\mathrm{E}[R^{i}] & =r_{k|k-1}^{i}(k)\mathrm{E}^{i}[R(x)]\label{eq:MeanR^i}
\end{align}
where $r_{k|k-1}^{i}(k)$ is given by (\ref{eq:CurrentTimeExistenceProb}),
and
\begin{align}
\mathrm{E}^{i}[h(x)] & =\int h(x)p_{k|k-1}^{i,k}(x)dx\label{eq:Expectationh(x)}\\
\mathrm{E}^{i}[R(x)] & =\int R(x)p_{k|k-1}^{i,k}(x)dx\\
\mathrm{E}^{i}[h(x)h(x)^{T}] & =\int h(x)h(x)^{T}p_{k|k-1}^{i,k}(x)dx.\label{eq:Expectationh(x)h(x)}
\end{align}
\end{lem}
\noindent A proof of the results in Lemma \ref{lem:Moments_of_y^i}
is provided in Appendix D. We next define the additive internal variables
across all targets except the $u$-th target, which we denote\foreignlanguage{british}{
\begin{align}
h^{(-u)} & =\sum_{i=1:i\neq u}^{n_{k|k-1}}h^{i}\label{eq:Sumh^i}\\
R^{(-u)} & =\sum_{i=1:i\neq u}^{n_{k|k-1}}R^{i}.\label{eq:SumR^i}
\end{align}
}Since the variables inside the sums in (\ref{eq:Sumh^i}) and (\ref{eq:SumR^i})
are independent, we directly obtain the following lemma.
\begin{lem}
\label{lem:The-mean-and}Given an MB predicted density of the form
(\ref{eq:TrajMBDensity}) with single-trajectory densities of the
form (\ref{eq:GaussianSingleTraj}), the mean and covariance of $h^{(-u)}$,
and the mean of $R^{(-u)}$, are
\begin{gather}
\hat{h}_{corr}^{u}=\mathrm{E}[h^{(-u)}]=\sum_{i=1:i\neq u}^{n_{k|k-1}}\mathrm{E}[h^{i}]\label{eq:MeanCorrection}\\
S_{corr}^{u}=\mathrm{C}[h^{(-u)}]=\sum_{i=1:i\neq u}^{n_{k|k-1}}\mathrm{C}[h^{i}]\label{eq:OtherTargetsCovar}\\
R_{corr}^{u}=\mathrm{E}[R^{(-u)}]=\sum_{i=1:i\neq u}^{n_{k|k-1}}\mathrm{E}[R^{i}]\label{eq:CovarCorrection}
\end{gather}
where $\mathrm{E}[h^{i}]$, $\mathrm{C}[h^{i}]$ and $\mathrm{E}[R^{i}]$
are given by (\ref{eq:Meany^i}), (\ref{eq:Covariancey^i}) and (\ref{eq:MeanR^i}),
respectively.
\end{lem}
To perform the update of the $u$-th Bernoulli, we require the conditional
moments $\mathrm{E}[z_{k}|\widetilde{\mathbf{x}}_{k}^{u}=\{(u,x^{u})\}]$
and $\mathrm{C}[z_{k}|\widetilde{\mathbf{x}}_{k}^{u}=\{(u,x^{u})\}]$
\cite{Tronarp2018}. For computational efficiency, we derive approximations
of these moments via Taylor series expansion of $m(\cdot)$ and $R(\cdot)$.
We do these expansions such that in the filter implementation we can
approximate the required integrals for each potential target by only
drawing sigma-points from the distribution of this potential target,
improving the computational efficiency of the algorithm.

Given the result of Lemma \ref{lem:The-mean-and}, the first-order
Taylor series expansion of $m(h(x^{u})+h^{(-u)})$ around the point
$h(x^{u})+\hat{h}_{corr}^{u}$ is given by
\begin{multline}
m(h(x^{u})+h^{(-u)})\approx m(h(x^{u})+\hat{h}_{corr}^{u})\\
+M(h(x^{u})+\hat{h}_{corr}^{u})(h^{(-u)}-\hat{h}_{corr}^{u})\label{eq:TaylorSeriesMean}
\end{multline}
where $M(h(x^{u})+\hat{h}_{corr}^{u})$ is the Jacobian of $m(\cdot)$
evaluated at $h(x^{u})+\hat{h}_{corr}^{u}$. In addition, the zero-order
Taylor series expansion of $\Sigma(R(x^{u})+R^{(-u)})$ around $R(x^{u})+R_{corr}^{u}$
is given by
\begin{equation}
\Sigma(R(x^{u})+R^{(-u)})\approx\Sigma(R(x^{u})+R_{corr}^{u}).\label{eq:TaylorSeriesCovar}
\end{equation}

\begin{prop}
\label{prop:ConditionalMoments}Under approximation (\ref{eq:TaylorSeriesMean}),
the conditional mean of $z_{k}$ given $\widetilde{\mathbf{x}}_{k}^{u}=\{(u,x^{u})\}$
is
\begin{equation}
\mathrm{E}[z_{k}|\widetilde{\mathbf{x}}_{k}^{u}=\{(u,x^{u})\}]\approx m(h(x^{u})+\hat{h}_{corr}^{u}).\label{eq:FullConditionalMean}
\end{equation}
Under approximations (\ref{eq:TaylorSeriesMean}) and (\ref{eq:TaylorSeriesCovar}),
the conditional covariance of $z_{k}$ given $\widetilde{\mathbf{x}}_{k}^{u}=\{(u,x^{u})\}$
is
\begin{multline}
\mathrm{C}[z_{k}|\widetilde{\mathbf{x}}_{k}^{u}=\{(u,x^{u})\}]\approx\Sigma(R(x^{u})+R_{corr}^{u})\\
+M(h(x^{u})+\hat{h}_{corr}^{u})S_{corr}^{u}M(h(x^{u})+\hat{h}_{corr}^{u})^{T}.\label{eq:FullConditionalCovar}
\end{multline}

For the case $\widetilde{\mathbf{x}}_{k}^{u}=\emptyset$, the conditional
moments are approximated as
\begin{gather}
\mathrm{E}[z_{k}|\widetilde{\mathbf{x}}_{k}^{u}=\emptyset]\approx m(\hat{h}_{corr}^{u})\label{eq:MeanNoTargetU}\\
\mathrm{C}[z_{k}|\widetilde{\mathbf{x}}_{k}^{u}=\emptyset]\approx\mathrm{\Sigma}\left(R_{corr}^{u}\right)+M(\hat{h}_{corr}^{u})S_{corr}^{u}M(\hat{h}_{corr}^{u})^{T}.\label{eq:CovarNoTargetU}
\end{gather}
\end{prop}
\noindent A proof of Proposition \ref{prop:ConditionalMoments} is
provided in Appendix E.

\selectlanguage{british}%
We clarify that using a first-order Taylor expansion for the conditional
mean and a zero-order Taylor expansion for the conditional covariance
is a standard type of approximation in Gaussian non-linear filtering
\cite[eq. (8)]{Tronarp2018}. In particular, this extends the standard
Extended Kalman filter (EKF) approximation, developed for Gaussian
measurement densities with a nonlinear conditional mean, to models
defined by conditional moments, keeping the EKF properties of tractability
and computational efficiency.
\selectlanguage{english}%

\subsection{\label{subsec:GaussianUpdate}Single-Target Gaussian Update}

\subsubsection{Enabling Approximation}

To perform a Gaussian update of the measurement model with conditional
moments (\ref{eq:FullConditionalMean}) and (\ref{eq:FullConditionalCovar}),
we must first approximate the non-linear and non-Gaussian measurement
into a linear and Gaussian form \cite{Arasaratnam2007}. We can view
the measurement (conditioned on a single-target $\widetilde{\mathbf{x}}_{k}^{u}=\{(u,x^{u})\}$)
as a transformation of the state $x^{u}$ by a non-linear function
$g(\cdot)$, where a state dependent zero-mean noise term $\eta(\cdot)$
with covariance matrix $\Psi(\cdot)$ is additive to this transformation.
This can be written as \cite{F.GarciaFernandez2019}
\begin{equation}
z_{k}=g(x^{u})+\eta(x^{u})\label{eq:GaussianMeasurementApproximation}
\end{equation}
where the non-linear function and noise covariance are given by
\begin{align}
g(x^{u}) & =\mathrm{E}[z_{k}|\widetilde{\mathbf{x}}_{k}^{u}=\{(u,x^{u})\}]\\
\Psi(x^{u}) & =\mathrm{C}[z_{k}|\widetilde{\mathbf{x}}_{k}^{u}=\{(u,x^{u})\}]\label{eq:Psix^u}
\end{align}
respectively, where $\mathrm{E}[z_{k}|\widetilde{\mathbf{x}}_{k}^{u}=\{(u,x^{u})\}]$
is given by (\ref{eq:FullConditionalMean}), and $\mathrm{C}[z_{k}|\widetilde{\mathbf{x}}_{k}^{u}=\{(u,x^{u})\}]$
is given by (\ref{eq:FullConditionalCovar}).

Considering the measurement as (\ref{eq:GaussianMeasurementApproximation}),
we form a linear and Gaussian approximation using the following enabling
approximation
\begin{equation}
z_{k}\approx Ax^{u}+b+r\label{eq:LinearGaussMeasurement}
\end{equation}
where $A\in\mathbb{R}^{Mn_{z}\times n_{x}}$, $b\in\mathbb{R}^{Mn_{z}}$
are linearisation parameters, and $r\in\mathbb{R}^{Mn_{z}}$ is a
zero-mean Gaussian noise term with covariance matrix $\Omega\in\mathbb{R}^{Mn_{z}\times Mn_{z}}$.
We clarify that $(A,b)$ provide an affine transformation which forms
a linear approximation to the conditional mean (\ref{eq:FullConditionalMean}),
and the Gaussian noise term $r$ has covariance matrix $\Omega$ which
accounts for both the linearisation error, and the conditional covariance
of the measurement (\ref{eq:FullConditionalCovar}). Given (\ref{eq:LinearGaussMeasurement}),
the KF provides us with the updated mean $\mu_{k|k}^{u}$ and covariance
$\Xi_{k|k}^{u}$ \cite{Saerkkae2023}
\begin{gather}
K^{u}=\Xi_{k|k-1}^{u}A^{T}(A\Xi_{k|k-1}^{u}A^{T}+\Omega)^{-1}\label{eq:KFUpdate1}\\
\mu_{k|k}^{u}=\mu_{k|k-1}^{u}+K^{u}(z_{k}-A\mu_{k|k-1}^{u}-b)\\
\Xi_{k|k}^{u}=\Xi_{k|k-1}^{u}-K^{u}(A\Xi_{k|k-1}^{u}A^{T}+\Omega)(K^{u})^{T}.\label{eq:KFUpdate5}
\end{gather}

Finally, the target probability of existence can be computed using
(\ref{eq:ProbabilityExistenceAlive}), which requires the likelihoods
of existence and non-existence. Under the Gaussian approximation of
the measurement (\ref{eq:GaussianMeasurementApproximation}) (with
moments defined in (\ref{eq:FullConditionalMean}) and (\ref{eq:FullConditionalCovar})),
the likelihood of existence is
\begin{equation}
p^{u}(z_{k}||\widetilde{\mathbf{x}}_{k}^{u}|=1)=\mathcal{N}(z_{k};A\mu_{k|k-1}^{u}+b,A\Xi_{k|k-1}^{u}A^{T}+\Omega).\label{eq:GaussianExistenceLikelihood}
\end{equation}
Under a Gaussian approximation of the measurement when the $u$-th
target is not present, using (\ref{eq:MeanNoTargetU}) and (\ref{eq:CovarNoTargetU}),
the likelihood of non-existence is
\begin{multline}
p^{u}(z_{k}|\emptyset)=\\
\mathcal{N}(z_{k};m(\hat{h}_{corr}^{u}),\Sigma(R_{corr}^{u})+M(\hat{h}_{corr}^{u})S_{corr}^{u}M(\hat{h}_{corr}^{u})^{T}).\label{eq:GaussianNonExistenceLikelihood}
\end{multline}
The parameters $(A,b,\Omega)$ are determined using Statistical Linear
Regression (SLR), which is explained in the next section.

\subsubsection{\label{subsec:SLRSigmaPoints}SLR using Sigma Points}

To perform a closed form Gaussian update, we require the linear and
Gaussian approximation to the measurement (\ref{eq:LinearGaussMeasurement}).
The required parameters $(A,b,\Omega)$ are found using SLR, which
returns an optimal linearisation of the nonlinear function $g(\cdot)$
with respect to some density $\mathcal{N}(\cdot;\overline{x},P)$.
More specifically, SLR minimises the mean squared error between the
linear and Gaussian approximation of the function, given by
\begin{align}
(A^{+},b^{+}) & =\underset{(A,b)}{\arg\min}\thinspace\mathrm{E}\left[||g(x^{u})-A^{+}x^{u}-b^{+}||^{2}\right]
\end{align}
where the zero-mean noise $\eta(x^{u})$ can be removed as this is
uncorrelated with $g(x^{u})$ \cite{Tronarp2018}. The solution is
\cite{Arasaratnam2007}
\begin{gather}
A^{+}=\mathrm{C}[x,g(x)]^{T}P^{-1}\label{eq:SLREq1}\\
b^{+}=\mathrm{E}[g(x)]-A^{+}\overline{x}^{u}.
\end{gather}
The resulting mean square error matrix is \cite{F.GarciaFernandez2019}
\begin{align}
\Omega & =\mathrm{C}[g(x^{u})]+\mathrm{E}[\Psi(x^{u})]-A^{+}P(A^{+})^{T}\label{eq:SLROmega}
\end{align}
where $\Psi(x^{u})$ was given in (\ref{eq:Psix^u}).

\cprotect\subsubsection{\label{subsec:Application-of-SLR}Application of SLR to IEMB Update
Step
\begin{algorithm}[t]
\protect\caption{\label{alg:SLRGeneralisedModel}SLR of Nested Superpositional Measurement
Model for the $u$-th Target.}

$\mathbf{Input:}$ $\bar{x}$, $P$, $\hat{h}_{corr}^{u}$, $R_{corr}^{u}$
and $S_{corr}^{u}$

$\mathbf{Output:}$ $(A,b,\Omega)$
\begin{itemize}
\item Draw $m_{s}$ sigma points $\mathcal{X}^{1},...,\mathcal{X}^{m_{s}}$
and corresponding weights which match $\bar{x}$, $P$ using an appropriate
method e.g., \cite[Chapter 8]{Saerkkae2023}.
\item Transform the set of sigma points through the nonlinear functions
in (\ref{eq:FullConditionalMean}) and (\ref{eq:FullConditionalCovar})
to
\begin{gather*}
\mathcal{G}^{s}=m(h(\mathcal{X}^{s})+\hat{h}_{corr}^{u}),\\
\mathcal{R}^{s}=\Sigma(R(\mathcal{X}^{s})+R_{corr}^{u}),\\
\mathcal{M}^{s}=M(h(\mathcal{X}^{s})+\hat{h}_{corr}^{u}).
\end{gather*}
\item The moments required to form $(A,b,\Omega)$ in (\ref{eq:SLREq1})-(\ref{eq:SLROmega})
are found using
\begin{gather*}
\mathrm{C}[x,g(x)]\approx\sum_{s=1}^{m_{s}}\omega^{s}(\mathcal{X}^{s}-\overline{x})(\mathcal{G}^{s}-\mathrm{E}[g(x)])^{T}\\
\mathrm{E}[g(x)]\approx\sum_{s=1}^{m_{s}}\omega^{s}\mathcal{G}^{s}\\
\mathrm{C}[g(x)]=\mathrm{E}[g(x)g(x)^{T}]-\mathrm{E}[g(x)]\mathrm{E}[g(x)]^{T}\\
\mathrm{E}[\Psi(x)]=\mathrm{E}[\Sigma(R(x)+R_{corr}^{u})]+\\
\mathrm{E}[M(h(x)+\hat{h}_{corr}^{u})S_{corr}^{u}M(h(x)+\hat{h}_{corr}^{u})^{T}]
\end{gather*}
where
\begin{gather*}
\mathrm{E}[g(x)g(x)^{T}]\approx\sum_{s=1}^{m_{s}}\omega^{s}\mathcal{G}^{s}(\mathcal{G}^{s})^{T}\\
\mathrm{E}[\Sigma(R(x)+R_{corr}^{u})]\approx\sum_{s=1}^{m_{s}}\omega^{s}\mathcal{R}^{s}\\
\mathrm{E}[M(h(x)+\hat{h}_{corr}^{u})S_{corr}^{u}M(h(x)+\hat{h}_{corr}^{u})^{T}]\approx\\
\sum_{s=1}^{m_{s}}\omega^{s}\mathcal{M}^{s}S_{corr}^{u}(\mathcal{M}^{s})^{T}.
\end{gather*}
\item Compute $(A,b,\Omega)$ using $\mathrm{E}[g(x)],\mathrm{C}[g(x)],\mathrm{C}[x,g(x)]$
and $\mathrm{E}[\Psi(x)]$ in (\ref{eq:SLREq1})-(\ref{eq:SLROmega}).
\end{itemize}
\end{algorithm}
}

The parameters $(A,b,\Omega)$ approximate the conditional moments
of the measurement (\ref{eq:FullConditionalMean}) and (\ref{eq:FullConditionalCovar}),
which contain the correction moments $\hat{h}_{corr}^{u}$, $R_{corr}^{u}$
and $S_{corr}^{u}$ that account for the influence of all other targets
on the measurement. Given this, we must firstly estimate the correction
moments before proceeding with SLR.

Based on (\ref{eq:FullConditionalMean}) and (\ref{eq:FullConditionalCovar}),
we adopt a sigma-point based approach for estimating both the correction
moments, and the moments required to perform SLR using (\ref{eq:SLREq1})-(\ref{eq:SLROmega}).
Alternative approaches could be employed, such as a Taylor series
expansion (as is used for performing SLR in \cite{F.GarciaFernandez2019}
and \cite{F.GarciaFernandez2021}), however we only detail an Unscented
Transform based \cite{Saerkkae2023} approach here.

To estimate the correction moments, we draw a set of $m_{s}=2n_{x}+1$
sigma-points denoted $\mathcal{X}^{1},...,\mathcal{X}^{m_{s}}$ with
corresponding weights $\omega^{1},...,\omega^{m_{s}}$, with respect
to each $i$-th predicted single-target density $\mathcal{N}(\cdot;\mu_{k|k-1}^{i},\Xi_{k|k-1}^{i})$.
The sigma-points and weights are drawn using an appropriate method
\cite[Chapter 5]{Saerkkae2023}, which are then used to approximate
the following moments
\begin{gather}
\mathrm{E}^{i}[h(x)]\approx\sum_{s=1}^{m_{s}}\omega^{s}h(\mathcal{X}^{s})\label{eq:CorrectionMoments1}\\
\mathrm{E}^{i}[h(x)h(x)^{T}]\approx\sum_{s=1}^{m_{s}}\omega^{s}h(\mathcal{X}^{s})(h(\mathcal{X}^{s}))^{T}\\
\mathrm{E}^{i}[R(x)]\approx\sum_{s=1}^{m_{s}}\omega^{s}R(\mathcal{X}^{s})\label{eq:CorrectionMoments2}
\end{gather}
for all potential targets. The resulting SLR using the conditional
moments (\ref{eq:FullConditionalMean}) and (\ref{eq:FullConditionalCovar})
with respect to a density with mean $\bar{x}$ and covariance matrix
$P$ is provided in Algorithm \ref{alg:SLRGeneralisedModel}.

\subsubsection{IPLF}

Executing Algorithm 1 with respect to $\mathcal{N}(\cdot;\mu_{k|k-1}^{u},\Xi_{k|k-1}^{u})$
returns $(A,b,\Omega)$ with respect to the prior, and performing
the update (\ref{eq:KFUpdate1})-(\ref{eq:KFUpdate5}) using this
parameter set is then equivalent to an Unscented KF (UKF) update.
An improved update can be performed using the IPLF \cite{GarciaFernandez2015},
which performs iterated SLR with respect to the best approximation
of the single-target posterior at each iteration.

Algorithm \ref{alg:IPLFSingleTargetUpdate} details the complete T-IEMB-IPLF
update stage for each single-target using the nested superpositional
measurement model.

\begin{algorithm}[t]
\caption{\label{alg:IPLFSingleTargetUpdate}T-IEMB-IPLF Update}

\textbf{Input}: Predicted Bernoulli components for $u=1:n_{k|k-1}$.

\textbf{Output}: Updated Bernoulli components for $u=1:n_{k|k}$.

\textbf{For $i=1:n_{k|k-1}$}
\begin{itemize}
\item Draw sigma points $\mathcal{X}^{1:m_{s}}$ and weights $\omega^{1:m_{s}}$
matching $\mu_{k|k-1}^{i}$ and $\Xi_{k|k-1}^{i}$.
\item Calculate $\mathrm{E}^{i}[h(x)]$, $\mathrm{E}^{i}[h(x)h(x)]$ and
$\mathrm{E}^{i}[R(x)]$ in (\ref{eq:CorrectionMoments1})-(\ref{eq:CorrectionMoments2})
using $\mathcal{X}^{1:m_{s}}$ and $\omega^{1:m_{s}}$.
\end{itemize}
\textbf{End}

\textbf{For $u=1:n_{k|k-1}$}
\begin{itemize}
\item Calculate $\hat{h}_{corr}^{u}$, $R_{corr}^{u}$ and $S_{corr}^{u}$
using (\ref{eq:MeanCorrection})-(\ref{eq:CovarCorrection}).
\end{itemize}
\textbf{End}

\textbf{For $u=1:n_{k|k-1}$}
\begin{itemize}
\item Set $\bar{x}=\mu_{k|k-1}^{u}$ and $P=\Xi_{k|k-1}^{u}$.

\textbf{Repeat}:
\begin{itemize}
\item Execute Algorithm 1 with inputs $\bar{x}$, $P$ and $\hat{h}_{corr}^{u}$,
$R_{corr}^{u}$, $S_{corr}^{u}$ to compute $(A,b,\Omega)$.
\item Calculate the updated mean $\mu_{k|k}^{u}$ and covariance $\Xi_{k|k}^{u}$
using $(A,b,\Omega)$ in (\ref{eq:KFUpdate1})-(\ref{eq:KFUpdate5}).
\item Set $\bar{x}=\mu_{k|k}^{u}$ and $P=\Xi_{k|k}^{u}$.
\end{itemize}
\textbf{Until}: Fixed number of iterations or KLD convergence \cite[Section IV-D]{GarciaFernandez2015}.
\item Compute $p^{u}(z_{k}||\widetilde{\mathbf{x}}_{k}^{u}|=1)$ and $p^{u}(z_{k}|\emptyset)$
using (\ref{eq:GaussianExistenceLikelihood}) and (\ref{eq:GaussianNonExistenceLikelihood}),
respectively. Then for alive trajectories, update $r_{k|k}^{u}$ using
(\ref{eq:ProbabilityExistenceAlive}), and for all trajectories update
$r_{k|k}^{u}$ using (\ref{eq:ProbabilityExistenceAll}) and $\beta_{k|k}^{u}(l)$
using (\ref{eq:UpdateBeta}).
\item Update mean and covariance of past trajectory states, see Appendix
F.
\end{itemize}
\textbf{End}
\end{algorithm}

\subsection{\label{subsec:Single-Trajectory-Gaussian}GT-IEMB Single-Trajectory
Gaussian Update}

\subsubsection{\label{subsec:Alive-TrajectoriesUpdate}Alive Trajectories}

With a predicted T-MB density of the form (\ref{eq:TrajMBDensity}),
we perform the update of the GT-IEMB using a state decomposition approach.
That is, we compute the single-target updated density $p_{k|k}^{i,k}(x)$
corresponding to the target state at time step $k$ of trajectory
density $p_{k|k}^{i,k}(X)$ using the update outlined in Section \ref{subsec:GaussianUpdate}.
This returns the updated single-target moments $\mu_{k|k}^{i}$ and
$\Xi_{k|k}^{i}$, which can then be used to update past trajectory
states using the approach detailed in Appendix F, see also \cite{Beutler2009}\cite{GarciaFernandez2022}.
Finally, the updated existence probability $r_{k|k}^{i}$ is computed
with (\ref{eq:ProbabilityExistenceAlive}), using the existence likelihoods
(\ref{eq:GaussianExistenceLikelihood}) and (\ref{eq:GaussianNonExistenceLikelihood}).

\subsubsection{All Trajectories}

Given a predicted T-MB density of the form (\ref{eq:TrajMBDensity}),
we obtain the updated single-target moments $\mu_{k|k}^{i}$ and $\Xi_{k|k}^{i}$
following the same approach as for the update of alive trajectories,
presented in Section \ref{subsec:Alive-TrajectoriesUpdate}. These
moments are then used to update the past states of the trajectory
following Appendix F. We note that, the updated trajectory density
$p_{k|k-1}^{i,l}(X)$ for $l\in\{t^{i},...,k-1\}$ is given by its
predicted trajectory density i.e. we do not perform any update, as
these represent trajectories which ended at a previous time step.
The updated existence probability $r_{k|k}^{i}$ is then found using
(\ref{eq:ProbabilityExistenceAll}), and the updated Beta parameters
$\beta_{k|k}^{i}(l)$ for $l\in\{t^{i},...,k\}$ are found using (\ref{eq:UpdateBeta}).

\subsection{\label{subsec:Practical-Considerations}Practical Considerations}

As the trajectory length increases, the update becomes increasingly
computationally expensive. As a practical approximation, which is
employed in the other trajectory based filters e.g., \cite{GarciaFernandez2020c,GarciaFernandez2020,GarciaFernandez2019},
we use the $L$-scan approximation which limits the update of a trajectory
to a window of $L$ time steps. That is, when performing the past
trajectory update using Appendix F, we update only the last $L$ time
steps, which limits computational cost of the trajectory update. We
also bound the size of the covariance matrix to $Ln_{x}\times Ln_{x}$,
which is consistent with an update of the last $L$ time steps.

For the set of alive trajectories, we discard Bernoulli components
whose updated existence probability $r_{k|k}^{i}$ is below some pruning
threshold $\Gamma_{p}$, which indicates there is very low probability
of the trajectory being alive. For the set of all trajectories, we
again discard Bernoulli components with updated existence probability
below $\Gamma_{p}$, however Bernoulli components once considered
alive will always have a high existence probability. Given this, $\beta_{k|k}^{i}(k)$
is used to indicate the probability that the trajectory is still alive
i.e. if $\beta_{k|k}^{i}(k)$ is very low then there is very high
probability the trajectory ended at a previous time step $l\in\{t^{i},...,k-1\}$.
Therefore, if $\beta_{k|k}^{i}(k)$ is less than some threshold $\Gamma_{a}$,
we set $\beta_{k|k}^{i}(k)=0$ which indicates that the trajectory
has died, and it is no longer updated (but is still held in the set
of all trajectories).

\subsection{\label{subsec:Estimation}Estimation}

Given the T-IEMB posterior (\ref{eq:MBApprox}) at any time step,
we use the following estimators to extract the corresponding sets
of alive and all trajectories \cite{GarciaFernandez2020}. For the
set of alive trajectories at time step $k$, each alive trajectory
is given from the T-IEMB posterior as $\{(t^{i},\bar{x}_{k|k}^{i}):r_{k|k}^{i}>\Gamma_{d}\}$,
meaning the set of alive trajectories comprises of each trajectory
with existence probability above the detection threshold $\Gamma_{d}$.

For the set of all trajectories at time step $k$, the set comprises
of each trajectory from the T-IEMB posterior given as $\{(t^{i},\bar{x}_{k|k}^{i}(l^{*})):r_{k|k}^{i}>\Gamma_{d},l^{*}=\arg\underset{l}{\max}\beta_{k|k}^{i}(l)\}$.
This gives each trajectory ending at the time step with the highest
value of $\beta_{k|k}^{i}(l)$, and also each trajectory has existence
probability greater than the threshold $\Gamma_{d}$.

\subsection{Discussion}

\selectlanguage{british}%
This section discusses the choice of the approximate integration methods
used in the proposed Gaussian T-IEMB implementation. First, the moments
of the internal variables $\mathrm{E}^{i}[h(x)]$, $\mathrm{E}^{i}[h(x)h(x)^{T}]$,
$\mathrm{E}^{i}[R(x)]$ are computed via sigma-point integration,
which is typically more accurate than analytical linearisation. Nevertheless,
to compute the conditional moments we use analytical linearisation,
see Proposition \ref{prop:ConditionalMoments}. The reason is that
developing sigma-point or quadrature-point integration methods for
this type of problem is not straightforward and can lead to a high
computational burden, as explained next.

In this case, a natural approach would be to vectorise the upper triangular
matrix of matrix $R^{i}$ to yield $\mathrm{vect}(R^{i})$ such that
we can compute its mean (already computed in matrix form) and also
its covariance matrix. Then, we would compute the mean and covariance
matrices of $h^{(-u)}$ and $\mathrm{vect}(R^{(-u)})$ using sigma
points. Now, to apply a sigma-point update, for each potential target,
we would need to stack the target state, $h^{(-u)}$ and $\mathrm{vect}(R^{(-u)})$
and draw sigma-points (or quadrature points) from this joint state.
This is a high-dimensional space (as typically there is a large number
of sensors) which would require a high number of sigma-points, making
the method computationally inefficient compared to analytical linearisation.
Sample-based approximations of the conditional moments are also possible
but this would imply a significant increase in computational burden,
which we aim to avoid.
\selectlanguage{english}%

\section{\label{sec:Simulations}Simulations}

We analyse the performance of the proposed algorithms by considering
a non-Gaussian TkBD scenario. We first detail the target dynamic model
and non-Gaussian measurements in Sections \ref{subsec:Target-Dynamic-Model}
and \ref{subsec:Rician-Distributed-Measurements}, respectively, before
detailing the considered filters and performance metric in Section
\ref{subsec:Performance-Assessment}. Simulation results and their
discussion are then provided in Section \ref{subsec:Simulation-Results}.

\subsection{\label{subsec:Target-Dynamic-Model}Target Dynamic Model}

\begin{figure}
\centering
\includegraphics[width=1\columnwidth]{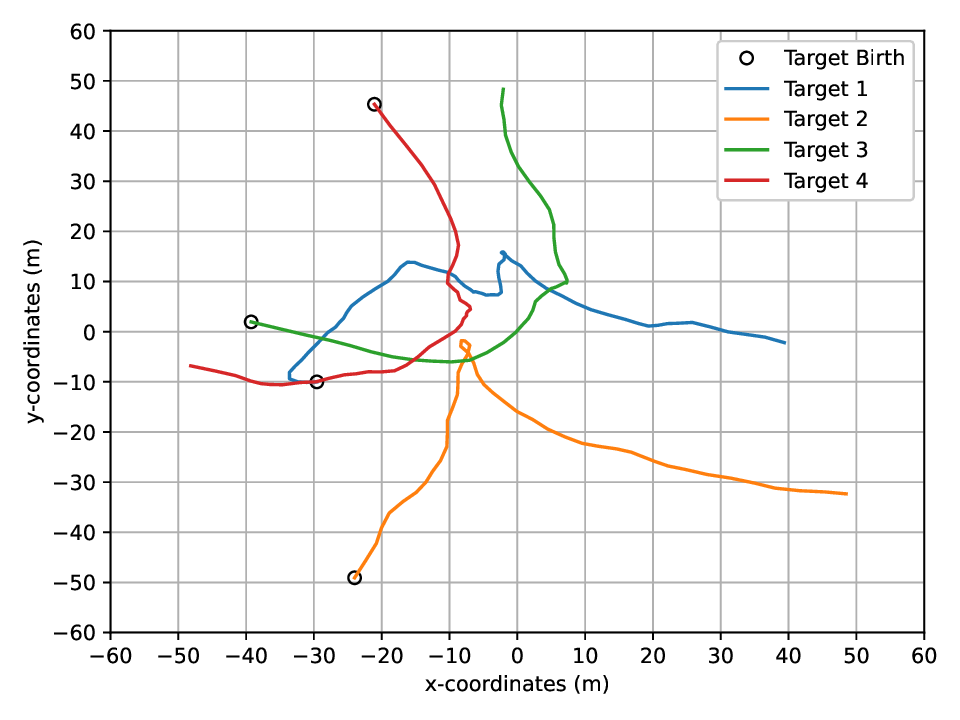}

\caption{\label{fig:Ground-truth-trajectories}Ground truth trajectories of
each target, where targets 1 through 4 are born at time steps 3, 16,
17, and 20, respectively, and die at time steps 74, 64, 57 and 64,
respectively. Each radar resolution cell is of dimensions $10\ \textrm{m}\times10\ \textrm{m}$,
with the perimeter of each cell illustrated using the grid.}
\end{figure}

A surveillance area of dimension $120\ \textrm{m}\times120\ \textrm{m}$
contains four targets, as illustrated in Figure \ref{fig:Ground-truth-trajectories}.
Each target is described by its state vector $x_{k}=[p_{x,k},v_{x,k},p_{y,k},v_{y,k}]^{T}$,
where $p_{x,k}$ and $v_{x,k}$ are the target position and velocity
in the x-axis at time step $k$, respectively. Each target survives
to the next time step with probability of survival $p^{S}=0.99$,
with state evolution between time steps modelled by a nearly constant
velocity model, where \cite{BarShalom2001}
\begin{equation}
F=I_{2}\otimes\left[\begin{array}{cc}
1 & T\\
0 & 1
\end{array}\right],\ Q=\sigma_{q}^{2}I_{2}\otimes\left[\begin{array}{cc}
\frac{T^{3}}{3} & \frac{T^{2}}{2}\\
\frac{T^{2}}{2} & T
\end{array}\right]
\end{equation}
are the state transition function and process noise covariance, respectively.
We note that $I_{2}$ denotes the two dimensional identity matrix,
$\otimes$ denotes the Kronecker product, $T=1\ $s is the sampling
period, and $\sigma_{q}^{2}=0.25\ m/s^{3/2}$ is the process noise
variance. We analyse two types of birth model: uninformative and informative,
whose details will be provided in Section \ref{subsec:Simulation-Results}.

\subsection{\label{subsec:Rician-Distributed-Measurements}Rician Distributed
Measurements}

We assume that each sensor measurement $z_{k}^{j}$ is independent
given the target state with the factorisation
\begin{equation}
p(z_{k}|\mathbf{x}_{k})=\prod_{j=1}^{M}p(z_{k}^{j}|\mathbf{x}_{k}).
\end{equation}
This implies that (\ref{eq:GeneralisedMean}) and (\ref{eq:GeneralisedCovariance})
can be defined for each sensor $j$, with inputs $h^{j}(\cdot)$,
$m^{j}(\cdot)$, $R^{j}(\cdot)$ and $\Sigma^{j}(\cdot)$ which consider
each sensor separately.

We consider a radar surveillance scenario, in which a surveillance
area is segmented into $M$ individual cells, where the centre of
cell $j\in\{1,...,M\}$ has coordinates $(c_{x}^{j},c_{y}^{j})$ in
the $x$ and $y$ dimensions, respectively. Each target in the surveillance
area is assumed to possess a constant RCS i.e. Swerling 0 targets,
and clutter is assumed to be Rayleigh distributed, which is a common
radar assumption. At each time step $k$, a radar return signal $z_{k}^{j}$
is measured from each cell, where the density of the measurement $z_{k}^{j}$
given the set of targets $\mathbf{x}_{k}$ is Rician of the form \cite[Chapter 8]{Mallick2013}
\begin{equation}
p(z_{k}^{j}|\mathbf{x}_{k})=\frac{z_{k}^{j}}{\sigma_{r}^{2}}\exp\left(-\frac{(z_{k}^{j})^{2}+(\lambda^{j}(\mathbf{x}_{k}))^{2}}{2\sigma_{r}^{2}}\right)I_{0}\left(\frac{z_{k}^{j}\lambda^{j}(\mathbf{x}_{k})}{\sigma_{r}^{2}}\right)\label{eq:RiceMeasurements}
\end{equation}
where $\sigma_{r}=2$, $\lambda^{j}(\mathbf{x}_{k})=\sum_{x_{k}\in\mathbf{x}_{k}}h^{j}(x_{k})$
where $h^{j}(x_{k})$ is the return signal from target $x_{k}$ in
cell $j$, and $I_{0}(\cdot)$ is the modified Bessel function of
the first kind of order $0$. The Rice density (\ref{eq:RiceMeasurements})
describes the measurement density of fixed RCS targets in the presence
of Rayleigh clutter \cite[Chapter 8]{Mallick2013}. Target returns
from each cell are modelled using the Gaussian Point Spread Function
(PSF)
\begin{equation}
h^{j}(x_{k})=\varphi\exp\left(-\frac{(c_{x}^{j}-p_{x,k})^{2}}{2\sigma_{x}^{2}}-\frac{(c_{y}^{j}-p_{y,k})^{2}}{2\sigma_{y}^{2}}\right)\label{eq:Riceh(x)}
\end{equation}
where $\varphi$ is the target amplitude, and $\sigma_{x}^{2}=\sigma_{y}^{2}=100$
is the variance of the PSF in each dimension (which defines the spatial
extent of the target signature).

Given the Rician distributed measurement (\ref{eq:RiceMeasurements}),
the conditional moments of each cell are\cite{Park1961}
\begin{align}
\mathrm{E}[z_{k}^{j}|\mathbf{x}_{k}] & =m^{j}(\lambda^{j}(\mathbf{x}_{k}))=\sigma_{r}\sqrt{\frac{\pi}{2}}L_{1/2}\left(-\frac{(\lambda^{j}(\mathbf{x}_{k})){}^{2}}{2\sigma_{r}^{2}}\right)\label{eq:RicianConditionalMean}\\
\mathrm{C}[z_{k}^{j}|\mathbf{x}_{k}] & =\Sigma^{j}(\lambda^{j}(\mathbf{x}_{k}))=2\sigma_{r}^{2}+(\lambda^{j}(\mathbf{x}_{k})){}^{2}-(\mathrm{E}[z_{k}^{j}|\mathbf{x}_{k}])^{2}
\end{align}
where $L_{1/2}(\cdot)$ denotes the Laguerre polynomial \cite[Chapter 22]{Abramowitz1972}.
We provide further information regarding the moments of the Rician
distributed measurements in Appendix G.

\subsection{\label{subsec:Performance-Assessment}Algorithms}

We consider two Gaussian implementations of the T-IEMB, the T-IEMB-UKF
and T-IEMB-IPLF. We also test the extension of the IEMB filter \cite{Davies2024}
for the nested measurement model, which is a side contribution of
this paper. This filter is obtained by marginalising out past trajectory
states in the T-IEMB implementation for alive trajectories \cite[Thm. 4]{Granstroem2025},
which reduces to applying the MB prediction step for sets of targets,
and not applying the past trajectory update in Appendix F after the
update on the set of targets. The estimated set of trajectories for
the IEMB filter is obtained via sequential track building (STB) by
linking target state estimates with the same auxiliary variable (track
index).

We also consider the Trajectory Independent Multi-Bernoulli (T-IMB)
filter which is an extension of the IMB filter \cite{Vo2010} to sets
of trajectories and the nested superpositional measurement model.
In other words, the T-IMB filter simply corresponds to the T-IEMB
filter without an exchange of information between filters. We have
tested UKF and IPLF implementations of the T-IMB filter. We cannot
compare with the IEMB filter in \cite{Davies2024} or the specific
IMB implementation in \cite{Vo2010}, since both filters were derived
for a standard superpositional sensor model, see Table \ref{tab:SensorModelProperties},
which cannot model (\ref{eq:RiceMeasurements}).

We compare against two multi-target TkBD particle filters. We consider
the Generalised Parallel Partition Multi-Bernoulli (GPP-MB) filter
\cite{GarciaFernandez2016}, and the Labelled Multi-Bernoulli for
the Generic Observation Model (LMB-GOM) filter \cite[Sec. IV]{Li2018}
with transition density proposal. Both filters were implemented using
$N_{p}$ particles (where GPP-MB uses $N_{p}$ particles per target
and LMB-GOM uses $N_{p}$ particles across all targets), and use a
sequential track-building approach to build tracks from their sequential
single-target estimates.

Both the UKF and IPLF use sigma-points with a central weight $\omega_{0}=\frac{1}{3}$,
and the IPLF is set to a maximum of 20 iterations with a KLD threshold
of $10^{-1}$. We report alive and all trajectories above the detection
threshold $\Gamma_{d}=0.5$. For the set of alive trajectories, we
prune trajectories with an updated existence probability below $\Gamma_{p}=0.01$,
and for the set of all trajectories we do not update trajectories
whose updated termination probability $\beta_{k|k}(k)$ is below $\Gamma_{a}=0.01$.
All filters were implemented in Python\footnote{T-IEMB code will be available after publication at https://github.com/SionLynch/GT-IEMB},
except the GPP-MB filter which used the Matlab/MEX implementation\footnote{GPP-MB code is available at https://github.com/Agarciafernandez/Track-before-detect-GPP-MB-filter}
from \cite{GarciaFernandez2016}.
\begin{figure}
\centering
\includegraphics[width=0.95\columnwidth]{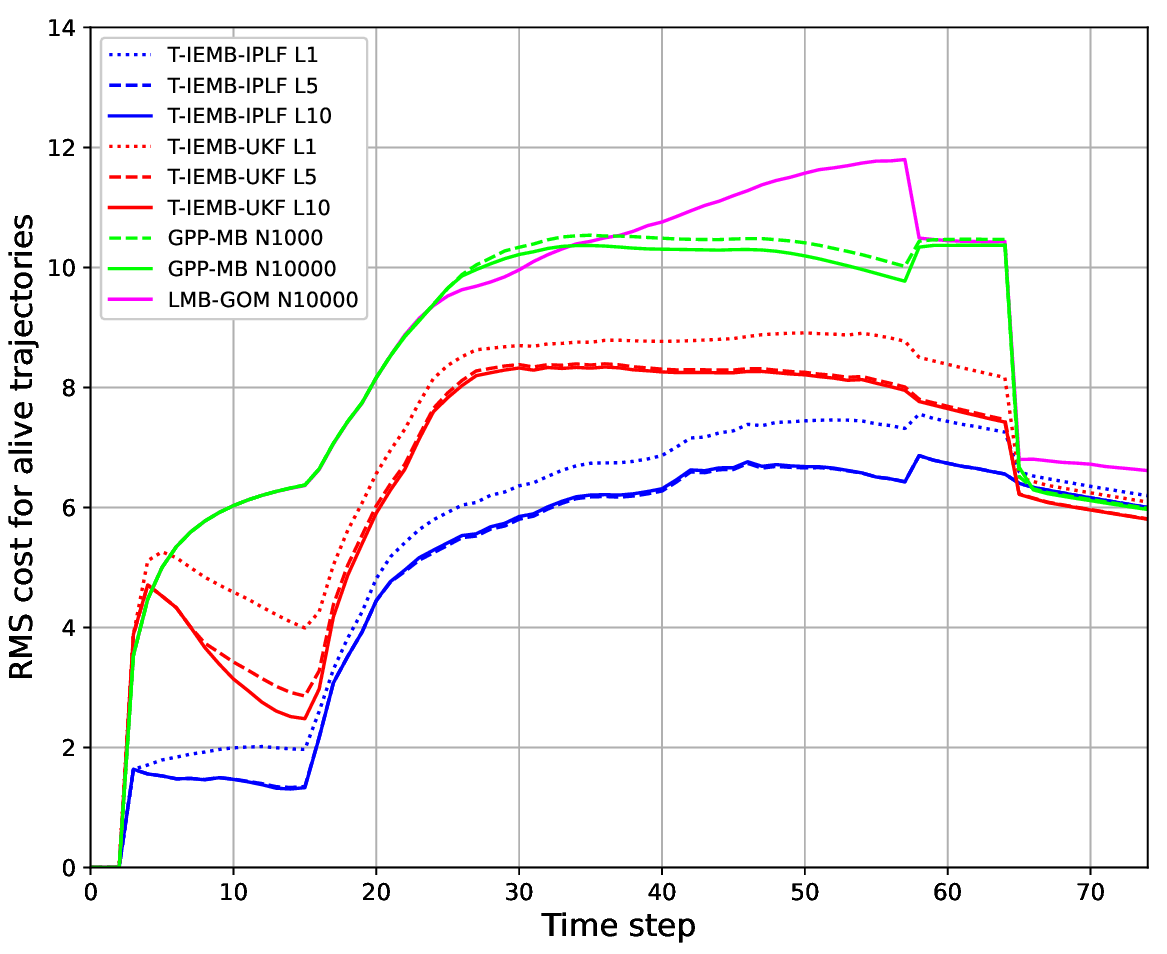}

\caption{\label{fig:TotalTGOSPA}Total RMS T-GOSPA cost of each filter for
the set of alive trajectories with uninformative birth. L$x$ denotes
a filter with an $L$-scan length of $x$, and N$y$ denotes a filter
with $y$ particles.}

\includegraphics[width=0.5\columnwidth]{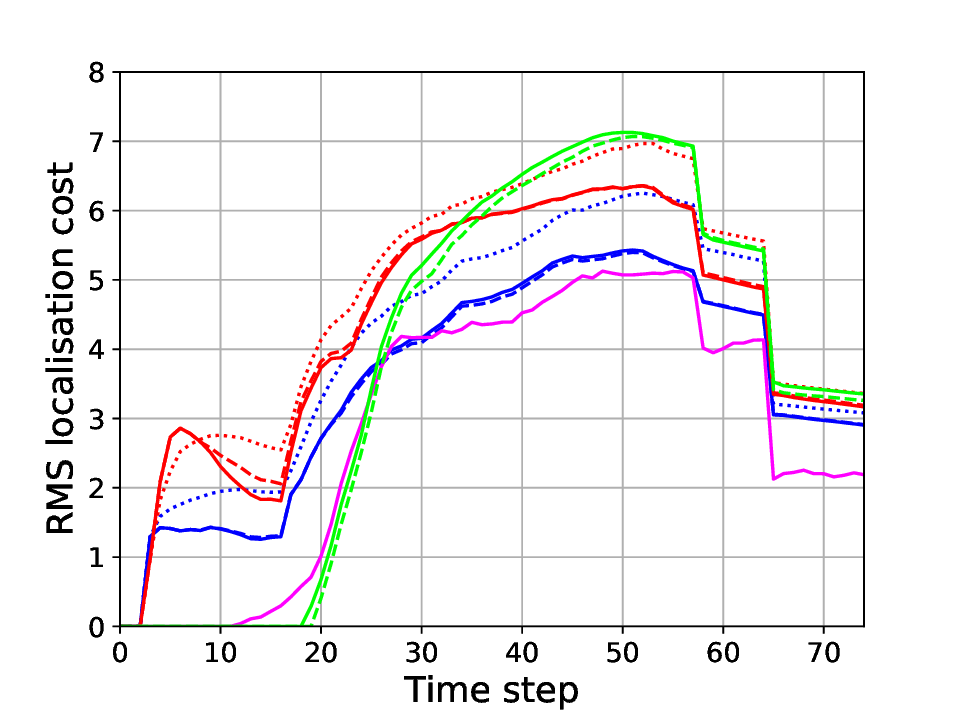}\includegraphics[width=0.5\columnwidth]{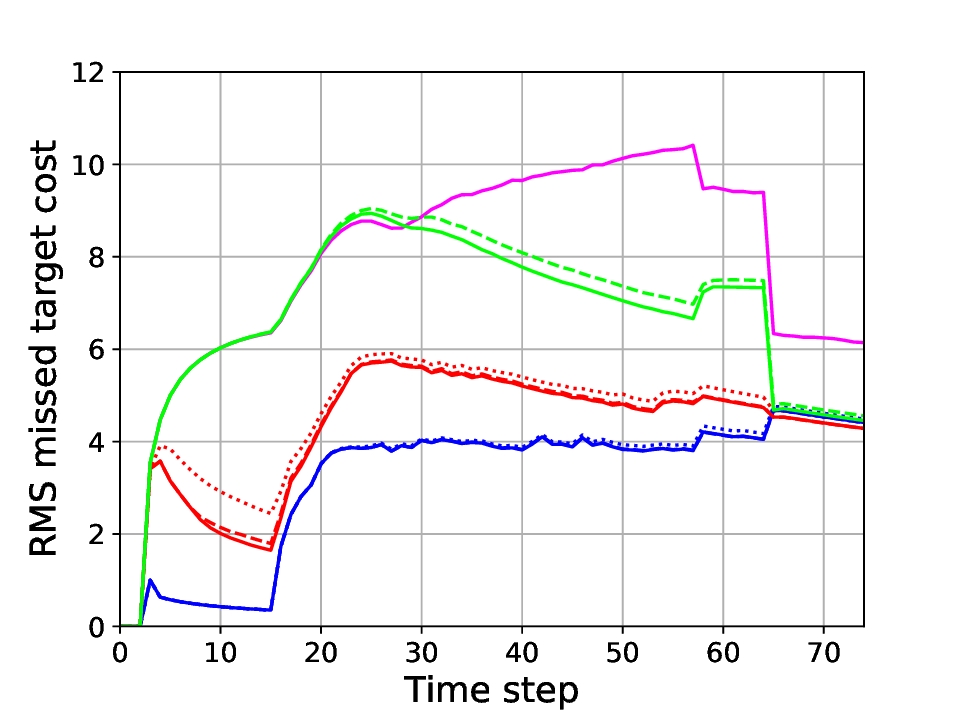}

\includegraphics[width=0.5\columnwidth]{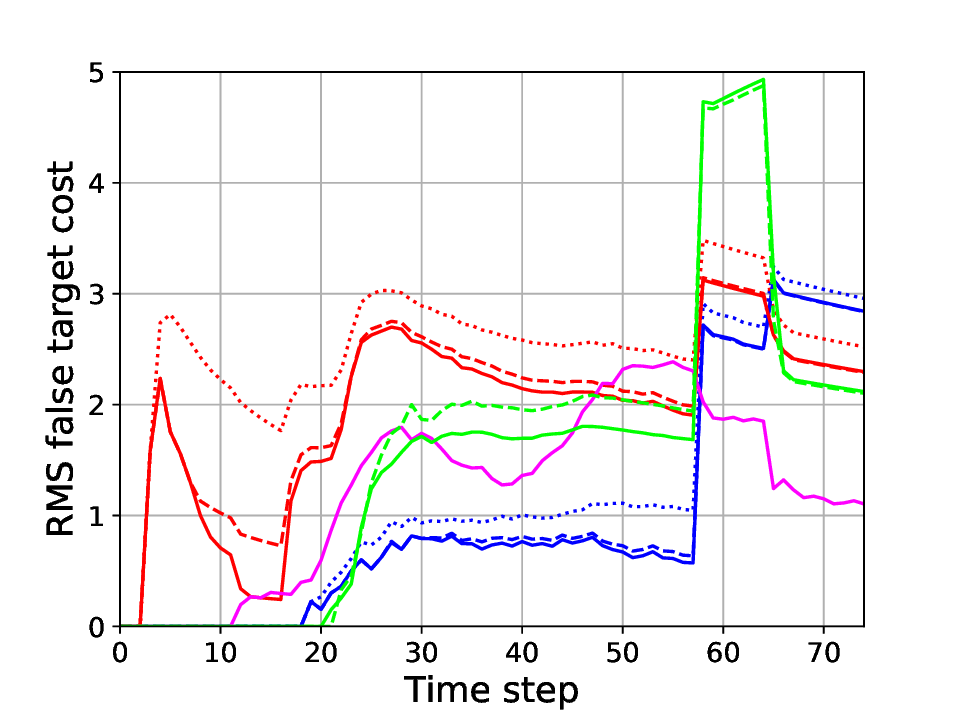}\includegraphics[width=0.5\columnwidth]{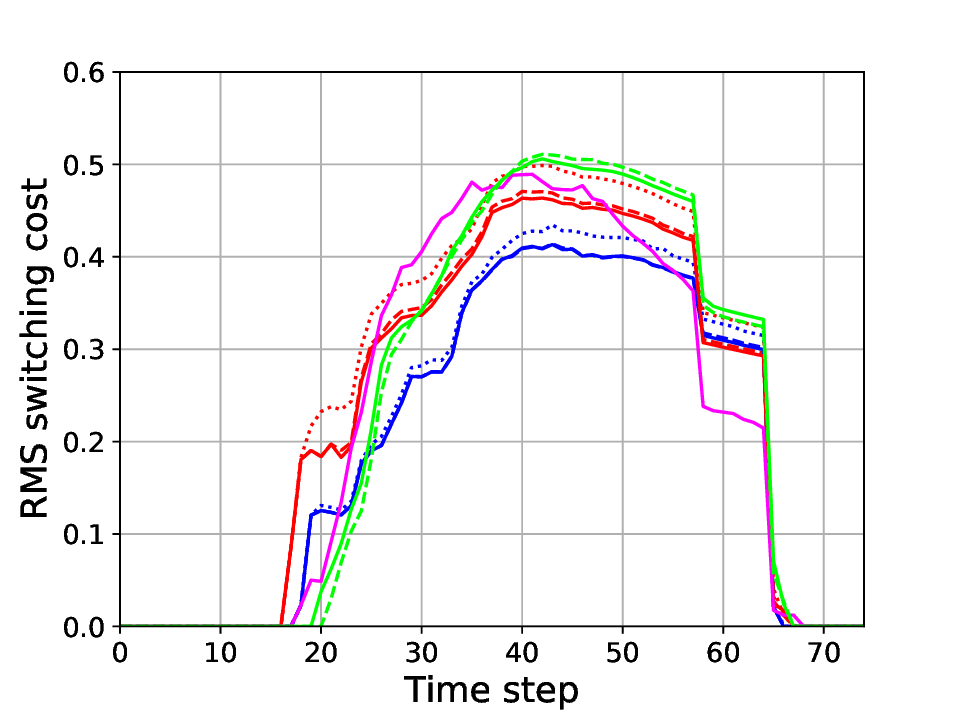}

\caption{\label{fig:DecomposedTGOSPA}(clockwise from top left) RMS T-GOSPA
localisation, missed, switch and false cost of each filter with uninformative
birth (legend is included in Figure \ref{fig:TotalTGOSPA}).}
\end{figure}
\begin{table*}
\centering
\caption{\label{tab:Table1}RMS T-GOSPA cost for alive trajectories with uninformative
birth}

\begin{tabular}{c|cc|c|c|c|c|c|ccc|ccc}
\hline 
 & \multicolumn{2}{c|}{GPP-MB} & LMB-GOM & \begin{cellvarwidth}[t]
\centering
STB-IEMB

-UKF
\end{cellvarwidth} & \begin{cellvarwidth}[t]
\centering
STB-IEMB

-IPLF
\end{cellvarwidth} & \begin{cellvarwidth}[t]
\centering
T-IMB

-UKF
\end{cellvarwidth} & \begin{cellvarwidth}[t]
\centering
T-IMB

-IPLF
\end{cellvarwidth} & \multicolumn{3}{c|}{T-IEMB-UKF} & \multicolumn{3}{c}{T-IEMB-IPLF}\tabularnewline
\hline 
$N_{p}$ & 1000 & 10000 & 10000 & - & - & - & - & - & - & - & - & - & -\tabularnewline
$L$-scan & - & - & - & - & - & 1 & 1 & 1 & 5 & 10 & 1 & 5 & 10\tabularnewline
\hline 
Total & 8.72 & 8.61 & 9.03 & 7.28 & 5.92 & 21.59 & 24.95 & 7.38 & 6.85 & 6.79 & 5.91 & 5.40 & 5.42\tabularnewline
Localisation & 4.61 & 4.71 & 3.45 & 5.01 & 4.42 & 5.06 & 5.26 & 5.04 & 4.68 & 4.65 & 4.43 & 3.83 & 3.85\tabularnewline
Missed & 7.08 & 6.91 & 8.21 & 4.75 & 3.63 & 5.70 & 6.01 & 4.71 & 4.50 & 4.47 & 3.58 & 3.53 & 3.53\tabularnewline
False & 2.11 & 2.04 & 1.45 & 2.29 & 1.54 & 20.18 & 23.62 & 2.58 & 2.16 & 2.10 & 1.56 & 1.43 & 1.42\tabularnewline
Switch & 0.31 & 0.32 & 0.30 & 0.32 & 0.27 & 0.84 & 0.70 & 0.32 & 0.30 & 0.30 & 0.27 & 0.26 & 0.26\tabularnewline
\hline 
\end{tabular}
\end{table*}
\begin{table}
\centering
\caption{\label{tab:RunTimeTable}Run time for each filter estimating alive
trajectories with uninformative birth}

\begin{tabular}{c|c|c}
\hline 
 & Parameter & Run time (s)\tabularnewline
\hline 
\multirow{2}{*}{GPP-MB} & $N_{p}=1000$ & 271.3\tabularnewline
 & $N_{p}=10000$ & 334.5\tabularnewline
\hline 
LMB-GOM & $N_{p}=10000$ & 3977.1\tabularnewline
\hline 
STB-IEMB-UKF & - & 1.6\tabularnewline
STB-IEMB-IPLF & - & 4.8\tabularnewline
\hline 
T-IMB-UKF & $L=1$ & 7.7\tabularnewline
T-IMB-IPLF & $L=1$ & 67.4\tabularnewline
\hline 
\multirow{2}{*}{T-IEMB-UKF} & $L=1$ & 1.6\tabularnewline
 & $L=10$ & 1.6\tabularnewline
\multirow{2}{*}{T-IEMB-IPLF} & $L=1$ & 4.8\tabularnewline
 & $L=10$ & 4.8\tabularnewline
\hline 
\end{tabular}
\end{table}

To assess filter performance we use the Trajectory Generalised Optimal
Sub-pattern Assignment (T-GOSPA) metric \cite{GarciaFernandez2020b},
denoted $d(\cdot,\cdot)$. In the T-GOSPA metric, assignments between
the trajectories in the ground truth and estimated trajectories can
change over time. When there is a change in assignment this is referred
to as a track switch. Track switches are more likely to occur when
targets move in close proximity. These assignments also determine
the concept of missed and false cost. T-GOSPA cost is decomposed into
four individual sources, namely localisation, missed, false and track
switch costs. The metric parameters used were $p=2$, $c=W$ where
$W$ is the radar resolution cell width, and $\gamma=1$.

We use an online T-GOSPA implementation for producing graphical results,
where we normalise the metric result at each time step $k$ considering
the set of trajectory estimates $\hat{\mathbf{X}}_{k}$ and true trajectories
$\mathbf{X}_{k}$. The Root Mean Squared (RMS) value of the metric
at each time step $k$ over $N_{mc}=100$ Monte Carlo (MC) runs is
then given by

\begin{equation}
d(k)=\sqrt{\frac{1}{N_{mc}k}\sum_{i=1}^{N_{mc}}d^{2}\left(\mathbf{X}_{k},\hat{\mathbf{X}}_{k}^{i}\right)}.
\end{equation}
We also generate tabulated results, where the RMS average metric result
over a scenario with $T$ time steps is given by
\begin{equation}
d_{T}=\sqrt{\frac{1}{T}\sum_{k=1}^{T}d^{2}(k)}.\label{eq:RMS_TGOSPA_Results}
\end{equation}

\subsection{\label{subsec:Simulation-Results}Simulation Results}

\subsubsection{\label{subsec:Uninformative-Birth}Uninformative Birth}

We first analyse the performance of the filters using an uninformative
birth model, meaning that targets may be born anywhere in the surveillance
area. We define $n_{k}^{b}=1$ Bernoulli birth component (\ref{eq:MBBirthComponents}),
where $p_{k}^{b,1}(\cdot)$ has mean and covariance $m_{k}^{b,1}=[0,0,0,0]^{T}$
and $P_{k}^{b,1}=\textrm{diag}([200,10,200,10])$, respectively, and
birth probability $r_{k}^{b,1}=10^{-6}$ (LMB-GOM required a higher
birth probability $r_{k}^{b,1}=10^{-2}$ to have similar performance).
Parameter $\varphi$ of the measurement model (\ref{eq:Riceh(x)})
is set to 10.

Figure \ref{fig:TotalTGOSPA} shows the total RMS T-GOSPA cost for
four of the considered filters when estimating the set of alive trajectories,
with varying $L$-scan lengths for each T-IEMB implementation. The
T-IMB and STB-IEMB filters are omitted from the figures for clarity,
with tabulated results provided instead. LMB-GOM performance is worst,
followed by GPP-MB, with the T-IEMB-UKF improving on the GPP-MB, and
the T-IEMB-IPLF providing the best performance. Increasing the number
of particles for GPP-MB provides some performance improvement, as
does increasing the $L$-scan length for the trajectory-based filters.
Figure \ref{fig:DecomposedTGOSPA} shows the T-GOSPA cost decomposition
for each filter considered in Figure \ref{fig:TotalTGOSPA}, which
allows for a more thorough analysis of each filters performance.

Localisation cost reduces when increasing the $L$-scan length for
both the UKF and IPLF implementation, as the update of past states
improves localisation over time, which demonstrates the benefit of
updating past trajectory states. Localisation is reduced for the IPLF
implementation as compared to the UKF, as its iterative update provides
a higher accuracy estimate of the target posterior density. The primary
performance difference between the T-IEMB implementations is in missed
and false target cost. The IPLFs iterative update means the highly
nonlinear update of a birth component can be performed over multiple
iterations, whereas the UKF must complete this update in a single
iteration. The IPLF implementation is therefore superior at initialising
new trajectories, which is evident from the reduced missed target
cost over the scenario. False target cost is also reduced for the
IPLF implementation, owing to an increased accuracy estimate of the
target posterior.

The increased cost for the GPP-MB is mainly influenced by a high missed
target cost, since it uses an uninformative birth density where birth
particles are spread across the surveillance area. This results in
it generally taking many time steps to initialise a new target, while
the birth densities for both T-IEMB implementations generally allow
for easier initialisation of trajectories.

LMB-GOM performance is also influenced by a high missed target cost,
partly due to the use of an uninformative birth density. The LMB-GOM
recursion involves summing over all subsets of the label space, which
grows exponentially with the number of labels. Given this, as the
number of targets increases the filter requires an increasing number
of particles to obtain good performance. This is evident as the missed
target cost increases throughout the scenario, as a result of incorrect
target pruning or not initialising new targets.
\begin{table*}
\centering
\caption{\label{tab:Table2}RMS T-GOSPA cost for all trajectories with uninformative
birth}

\begin{tabular}{c|c|c|ccc|ccc}
\hline 
 & STB-IEMB-UKF & STB-IEMB-IPLF & \multicolumn{3}{c|}{T-IEMB-UKF} & \multicolumn{3}{c}{T-IEMB-IPLF}\tabularnewline
\hline 
$L$-scan & - & - & 1 & 5 & 10 & 1 & 5 & 10\tabularnewline
\hline 
Total & 7.47 & 5.82 & 7.56 & 6.95 & 6.88 & 5.77 & 5.10 & 5.11\tabularnewline
Localisation & 5.57 & 4.93 & 5.61 & 5.17 & 5.14 & 4.94 & 4.24 & 4.26\tabularnewline
Missed & 4.43 & 2.90 & 4.37 & 4.13 & 4.09 & 2.81 & 2.72 & 2.72\tabularnewline
False & 2.24 & 1.00 & 2.55 & 2.12 & 2.04 & 0.98 & 0.71 & 0.68\tabularnewline
Switch & 0.38 & 0.32 & 0.38 & 0.35 & 0.35 & 0.32 & 0.31 & 0.31\tabularnewline
\hline 
\end{tabular}
\end{table*}
\begin{table*}
\centering
\caption{\label{tab:RobustnessAnalysisTable}Total RMS T-GOSPA cost for alive
trajectories with varying target amplitude and uninformative birth}

\begin{tabular}{c|c|c|c|c|c|c|cc|cc}
\hline 
\multicolumn{1}{c}{} &  & GPP-MB & STB-IEMB-UKF & STB-IEMB-IPLF & T-IMB-UKF & T-IMB-IPLF & \multicolumn{2}{c|}{T-IEMB-UKF} & \multicolumn{2}{c}{T-IEMB-IPLF}\tabularnewline
\hline 
\multicolumn{2}{c|}{$N_{p}$} & 1000 & - & - & - & - & - & - & - & -\tabularnewline
\multicolumn{2}{c|}{$L$-scan} & - & - & - & 1 & 1 & 1 & 10 & 1 & 10\tabularnewline
\hline 
\multirow{4}{*}{$\varphi$} & 6 & 9.06 & 8.76 & 7.94 & 16.02 & 18.27 & 8.82 & 8.22 & 7.96 & 7.38\tabularnewline
 & 8 & 8.87 & 7.85 & 6.52 & 18.52 & 23.14 & 7.96 & 7.35 & 6.49 & 5.94\tabularnewline
 & 10 & 8.72 & 7.28 & 5.92 & 21.59 & 24.95 & 7.38 & 6.79 & 5.91 & 5.42\tabularnewline
 & 12 & 8.69 & 7.32 & 5.89 & 23.05 & 25.00 & 7.36 & 6.82 & 5.89 & 5.50\tabularnewline
\hline 
\end{tabular}
\end{table*}

Each considered filter has a similar track switch cost, it being highest
around the midpoint of the scenario, since targets are in close proximity
and trajectory estimates are more likely to switch which ground truth
target they are estimating. The increase in missed and false target
costs coinciding with target deaths, around time steps 57 and 64,
is in fact a consequence of track switching. That is, since we are
estimating only alive trajectories, when a trajectory estimate switches
the portion before the track switch is then estimating a different
ground truth trajectory, which may die before the one it has switched
to. This results in a potential increase in missed and false target
cost at some future point, as observed in Figure \ref{fig:DecomposedTGOSPA}.
Such behaviour would not be observed when estimating the set of all
trajectories.

Table \ref{tab:Table1} contains the tabulated RMS T-GOSPA costs of
each filter estimating the set of alive trajectories, which were generated
using (\ref{eq:RMS_TGOSPA_Results}). The T-IEMB-IPLF is best performing,
with T-IMB-IPLF worst performing. Both STB-IEMB implementations show
similar performance to the corresponding T-IEMB. STB-IEMB-UKF is slightly
better performing than T-IEMB-UKF due to poor localisation and low
existence probability on target/trajectory initialisation. In this
case the STB does not report the first estimate as it is below the
detection threshold, which lowers its T-GOSPA cost. T-IEMB-IPLF is
slightly better performing than STB-IEMB-IPLF due to improved initialisation
of targets/trajectories.

Both T-IMB implementations are affected by a high false target cost,
as the IMB assumes target contributions do not overlap. When this
assumption does not hold, false targets are subsequently initialised.
We only include T-IMB results for $L=1$ since accurate estimation
of the number of targets is their primary performance issue, meaning
improving localisation does not provide much benefit. Further illustrated
is the reduction in cost by increasing the $L$-scan length for trajectory-based
filters.

Table \ref{tab:RunTimeTable} presents the run time for a single MC
run of each filter\footnote{GPP-MB was implemented using MATLAB/MEX, meaning its run time is not directly comparable to the Python implementations, providing a coarse comparison.}
for estimating the set of alive trajectories with uninformative birth.
All filters were run on a laptop with a 2.4 GHz Intel i5 CPU and integrated
Intel Iris Xe GPU. Both T-IEMB run times are low owing to their computationally
light Gaussian filtering solution, and accurate estimation of the
number of targets in the scenario. Increasing the $L$-scan length
from $L=1$ to $L=10$ has negligible effect on run time, since the
update of the past trajectory state (reviewed in Appendix F) has a
low computational load compared to the rest of the filter algorithm
(which is independent of the choice of $L$). While the T-IMB is also
a Gaussian filter, its very high false target cost increases its computational
load as the filter initialises and updates a high number of trajectories.

Both the GPP-MB and LMB-GOM have high run times due to their particle-based
approach. Increasing the number of particles for GPP-MB also increases
its run time (where the run time increase is not linear as computations
involving particles are implemented in MEX). LMB-GOM run time is very
high due to the exponential complexity of the algorithm in the number
of targets (where a grouping-based LMB-GOM has previously been proposed
\cite[Sec. IV-D]{Li2018} to address this when targets are well spaced).
Increasing the number of particles for LMB-GOM could reduce its T-GOSPA
cost, but would further increase run time.

Table \ref{tab:Table2} contains the tabulated RMS T-GOSPA costs for
each STB-IEMB and T-IEMB implementation for estimating the set of
all trajectories. T-IEMB-IPLF is again the best performing filter,
with increasing $L$-scan lengths also improving performance for each
T-IEMB implementation.

\subsubsection{Robustness Analysis}

To analyse the robustness of the proposed filters, we test each filter
using the parameter values $\varphi\in\{6,8,10,12\}$, and the uninformative
birth model presented in Section \ref{subsec:Uninformative-Birth}.
The total T-GOSPA cost for each filter estimating the set of alive
trajectories are presented in Table \ref{tab:RobustnessAnalysisTable},
where the LMB-GOM was omitted due to its high run time.

Performance for all filters (excluding T-IMB) is worst for the lowest
amplitude, as would be expected since it is difficult to distinguish
targets from noise. As the target amplitude increases, the T-GOSPA
cost for each filter generally decreases as would also be expected.
GPP-MB performance also improves as the target amplitude increases;
however not at the same rate as the T-IEMB filters. As previously
discussed, the uninformative birth model is still the primary performance
issue for the GPP-MB. T-IMB performance worsens as the target amplitude
increases, as the filter incorrectly initialises more trajectories
when the target amplitude is higher. STB-IEMB and T-IEMB performance
is again very similar.

\subsubsection{Informative Birth}

We now consider an informative birth model, meaning there is accurate
prior information on target birth locations.
\begin{table*}
\centering
\caption{\label{tab:InformativeBirthResults}RMS T-GOSPA cost for alive trajectories
with informative birth}

\begin{tabular}{c|cc|c|c|c|c|c|ccc|ccc}
\hline 
 & \multicolumn{2}{c|}{GPP-MB} & LMB-GOM & \begin{cellvarwidth}[t]
\centering
STB-IEMB

-UKF
\end{cellvarwidth} & \begin{cellvarwidth}[t]
\centering
STB-IEMB

-IPLF
\end{cellvarwidth} & \begin{cellvarwidth}[t]
\centering
T-IMB

-UKF
\end{cellvarwidth} & \begin{cellvarwidth}[t]
\centering
T-IMB

-IPLF
\end{cellvarwidth} & \multicolumn{3}{c|}{T-IEMB-UKF} & \multicolumn{3}{c}{T-IEMB-IPLF}\tabularnewline
\hline 
$N_{p}$ & 1000 & 10000 & 2500 & - & - & - & - & - & - & - & - & - & -\tabularnewline
$L$-scan & - & - & - & - & - & 1 & 1 & 1 & 5 & 10 & 1 & 5 & 10\tabularnewline
\hline 
Total & 5.86 & 5.69 & 9.44 & 5.25 & 5.18 & 26.39 & 29.42 & 5.25 & 4.60 & 4.61 & 5.18 & 4.56 & 4.56\tabularnewline
Localisation & 4.75 & 4.51 & 2.73 & 4.46 & 4.43 & 4.92 & 5.40 & 4.47 & 3.79 & 3.81 & 4.44 & 3.79 & 3.79\tabularnewline
Missed & 2.61 & 2.64 & 8.76 & 2.24 & 2.15 & 4.15 & 4.15 & 2.21 & 2.12 & 2.11 & 2.12 & 2.04 & 2.04\tabularnewline
False & 2.21 & 2.22 & 2.22 & 1.61 & 1.59 & 25.57 & 28.61 & 1.61 & 1.49 & 1.48 & 1.59 & 1.48 & 1.48\tabularnewline
Switch & 0.26 & 0.26 & 0.16 & 0.26 & 0.27 & 0.95 & 0.92 & 0.26 & 0.26 & 0.26 & 0.27 & 0.26 & 0.26\tabularnewline
\hline 
\end{tabular}
\end{table*}
\begin{table}
\centering
\caption{\label{tab:RunTimeTable-1}Run time for each filter estimating alive
trajectories with informative birth}

\begin{tabular}{c|c|c}
\hline 
 & Parameter & Run time (s)\tabularnewline
\hline 
\multirow{2}{*}{GPP-MB} & $N_{p}=1000$ & 435.5\tabularnewline
 & $N_{p}=10000$ & 469.2\tabularnewline
\hline 
LMB-GOM & $N_{p}=2500$ & 2090.7\tabularnewline
\hline 
STB-IEMB-UKF & - & 2.6\tabularnewline
STB-IEMB-IPLF & - & 6.6\tabularnewline
\hline 
T-IMB-UKF & $L=1$ & 12.5\tabularnewline
T-IMB-IPLF & $L=1$ & 106.7\tabularnewline
\hline 
\multirow{2}{*}{T-IEMB-UKF} & $L=1$ & 2.6\tabularnewline
 & $L=10$ & 2.6\tabularnewline
\multirow{2}{*}{T-IEMB-IPLF} & $L=1$ & 6.6\tabularnewline
 & $L=10$ & 6.6\tabularnewline
\hline 
\end{tabular}
\end{table}
We define $n_{k}^{b}=3$ Bernoulli birth components (\ref{eq:MBBirthComponents}),
where each $p_{k}^{b,i}(\cdot)$ for $i\in\{1,2,3\}$ has mean $m_{k}^{b,1}=[-20,0,45,0]^{T}$,
$m_{k}^{b,2}=[-35,0,-5,0]^{T}$ and $m_{k}^{b,3}=[-25,0,-50,0]^{T}$,
with covariance $P_{k}^{b,i}=\textrm{diag}([10,10,10,10])$, each
located in the area where a real target is born. We consider 3 birth
locations instead of 4 to lower the computational complexity, and
a birth probability $r_{k}^{b,i}=10^{-6}$, though LMB-GOM used $r_{k}^{b,i}=10^{-2}$
for improved performance. Parameter $\varphi$ of the measurement
model (\ref{eq:Riceh(x)}) is set to 10. For this experiment, the
LMB-GOM filter uses $N_{p}=2500$ since the run time for our implementation
with $N_{p}=10000$ was too large due to the exponential complexity
in the number of potential targets (which is increased in this scenario
due to the higher number of Bernoulli components in the birth process).

Table \ref{tab:InformativeBirthResults} contains the tabulated RMS
T-GOSPA costs for each filter for estimating the set of alive trajectories.
As in the previous scenario, T-IEMB-IPLF is best followed by the STB-IEMB-IPLF.
These filters are followed by the T-IEMB-UKF and STB-IEMB-UKF, then
GPP-MB, LMB-GOM, T-IMB-UKF and finally the T-IMB-IPLF. LMB-GOM performance
is mainly affected by targets being incorrectly pruned, which cannot
be re-initialised during the scenario since birth densities are only
at target birth locations. Performance could be improved by increasing
the number of particles. Table \ref{tab:RunTimeTable-1} presents
the run time for each filter.

\section{\label{sec:Conclusion}Conclusion}

In this work, we have extended the IEMB filter to estimate sets of
trajectories from nested superpositional measurements, which can be
used to model a broad range of TkBD measurements. The T-IEMB recursion
has been presented for estimating both the set of alive and all trajectories.
A Gaussian implementation has been proposed, which performs the update
using the conditional moments of the measurement. A Taylor series
linearisation was proposed to form an approximation of the conditional
moments of the nested superpositional measurements.

Simulation results for a non-Gaussian multi-target tracking scenario
have demonstrated the performance of two Gaussian implementations
of the T-IEMB filter, one using the UKF and another using the IPLF.
Both implementations outperformed the GPP-MB and LMB-GOM filters,
which are state-of-the-art particle filters for performing TkBD. Each
implementation of the T-IEMB filter also incurs a lower computational
cost than the particle filters, owing to the computationally light
structure of their Gaussian implementation.

Future work includes the application of the developed T-IEMB filter
to challenging non-Gaussian tracking scenarios e.g., K-distributed
sea clutter \cite{Ward2013}, and validation of the filters using
real data. Alternative approaches to forming the conditional moments
in Proposition \ref{prop:ConditionalMoments}, such as those relying
only on sigma-point and sample-based approaches, are also of interest.
Developing SMC implementations of the T-IEMB filters is another line
of future work.

\bibliographystyle{IEEEtran}
\bibliography{MTT}
\clearpage{}

{\Large A Track-Before-Detect Trajectory Multi-Bernoulli Filter for
Nested Superpositional Measurements: Supplemental Material}{\Large\par}

\appendices{}

\section{\label{subsec:Integrals-Over-Sets}}

This appendix reviews integration for functions on sets of trajectories.

Given some real valued function $\pi(\cdot)$ on the single-trajectory
space $T_{(k)}$, its integral is given by \cite{GarciaFernandez2020a}
\begin{equation}
\int\pi(X)dX=\sum_{(t,v)\in I_{(k)}}\int\pi(t,x^{1:v})dx^{1:v}
\end{equation}
where the summation goes through all possible start times, lengths
and target states. Given a real valued function $\pi(\cdot)$ on the
space of sets of trajectories $\mathcal{F}(T_{(k)})$, the set integral
is
\begin{equation}
\int\pi(\mathbf{X})\delta\mathbf{X}=\sum_{n=0}^{\infty}\frac{1}{n!}\int\pi(\{X_{1},...,X_{n}\})dX_{1:n}.
\end{equation}

\section{\label{subsec:Derivation-of-}}

This appendix derives the update for the set of all trajectories in
Lemma \ref{lem:-Given-a}.

\selectlanguage{british}%
Given the single-trajectory density form (\ref{eq:GeneralTrajForm}),
the posterior density for all trajectories is found by taking the
predicted Bernoulli of the form (\ref{eq:AuxiliaryBernoulli}) and
plugging into the update (\ref{eq:MBApproxComponent}). For $\widetilde{\mathbf{X}}_{k}=\{(u,X)\}$,
the posterior is
\begin{align}
\widetilde{f}_{k|k}^{u}(\{(u,X)\}) & \propto p^{u}(z_{k}|\tau^{k}(\{(u,X)\}))\tilde{f}_{k|k-1}(\{(u,X)\})\nonumber \\
 & =p^{u}(z_{k}|\tau^{k}(X))r_{k|k-1}^{u}\nonumber \\
 & \times\sum_{l=t^{u}}^{k}\beta_{k|k-1}^{u}(l)p_{k|k-1}^{u,l}(X)\nonumber \\
 & =r_{k|k-1}^{u}\left[p^{u}(z_{k}|\tau^{k}(X))\beta_{k|k-1}^{u}(k)p_{k|k-1}^{u,k}(X)\right.\nonumber \\
 & \quad\left.+p^{u}(z_{k}|\tau^{k}(X))\sum_{l=t^{u}}^{k-1}\beta_{k|k-1}^{u}(l)p_{k|k-1}^{u,l}(X)\right].
\end{align}
For $l\in\{t^{u},...,k-1\}$, we have that $p_{k|k-1}^{u,l}(X)=0$
if $\tau^{k}(X)\neq\emptyset$, since the density ends at time step
$l$. Therefore, we can write
\begin{align}
\widetilde{f}_{k|k}^{u}(\{(u,X)\}) & \propto r_{k|k-1}^{u}\left[p^{u}(z_{k}|\tau^{k}(X))\beta_{k|k-1}^{u}(k)p_{k|k-1}^{u,k}(X)\right.\nonumber \\
 & \quad\left.+p^{u}(z_{k}|\emptyset)\sum_{l=t^{u}}^{k-1}\beta_{k|k-1}^{u}(l)p_{k|k-1}^{u,l}(X)\right]\label{eq:AllTrajTargetPosterior}
\end{align}
For $\mathbf{\widetilde{X}}_{k}=\emptyset$, the posterior is
\begin{align}
\widetilde{f}_{k|k}^{u}(\emptyset) & \propto p^{u}(z_{k}|\emptyset)\tilde{f}_{k|k-1}(\emptyset)\nonumber \\
 & =p^{u}(z_{k}|\emptyset)(1-r_{k|k-1}^{u})\label{eq:AllTrajNoTargetPosterior}
\end{align}

\selectlanguage{english}%
Using the results (\ref{eq:AllTrajTargetPosterior}) and (\ref{eq:AllTrajNoTargetPosterior}),
we can write
\begin{multline}
\widetilde{f}_{k|k}^{u}(\widetilde{\mathbf{X}}_{k})\propto\begin{cases}
r_{k|k-1}^{u}\vartheta(X) & \widetilde{\mathbf{X}}_{k}=\{(u,X)\}\\
p^{u}(z_{k}|\emptyset)(1-r_{k|k-1}^{u}) & \widetilde{\mathbf{X}}_{k}=\emptyset\\
0 & \textrm{otherwise}
\end{cases}
\end{multline}
where
\begin{multline}
\vartheta(X)=p^{u}(z_{k}|\tau^{k}(X))\beta_{k|k-1}^{u}(k)p_{k|k-1}^{u,k}(X)\\
+p^{u}(z_{k}|\emptyset)\sum_{l=t^{u}}^{k-1}\beta_{k|k-1}^{u}(l)p_{k|k-1}^{u,l}(X).
\end{multline}
Let us simplify $\vartheta(X)$ as
\begin{alignat}{1}
\vartheta(X) & =\beta_{k|k-1}^{u}(k)\left[\int p^{u}(z_{k}|\tau^{k}(X))p_{k|k-1}^{u,k}(X)dX\right]\nonumber \\
 & \times\frac{p^{u}(z_{k}|\tau^{k}(X))p_{k|k-1}^{u,k}(X)}{\int p^{u}(z_{k}|\tau^{k}(X))p_{k|k-1}^{u,k}(X)dX}\nonumber \\
 & +p^{u}(z_{k}|\emptyset)\sum_{l=t^{u}}^{k-1}\beta_{k|k-1}^{u}(l)p_{k|k-1}^{u,l}(X)\nonumber \\
 & =\beta_{k|k-1}^{u}(k)p^{u}(z_{k}||\mathbf{\widetilde{x}}_{k}|=1)p_{k|k}^{u,k}(X)\nonumber \\
 & +p^{u}(z_{k}|\emptyset)\sum_{l=t^{u}}^{k-1}\beta_{k|k-1}^{u}(l)p_{k|k-1}^{u,l}(X)\label{eq:ThetaInitial}
\end{alignat}
where
\begin{gather}
p^{u}(z_{k}||\mathbf{\widetilde{x}}_{k}|=1)=\int p^{u}(z_{k}|\tau^{k}(X))p_{k|k-1}^{u,k}(X)dX\\
p_{k|k}^{u,l}(X)=\begin{cases}
p_{k|k-1}^{u,l}(X) & l\in\{t^{u},...,k-1\}\\
\frac{p^{u}(z_{k}|\tau^{k}(X))p_{k|k-1}^{u,k}(X)}{\int p^{u}(z_{k}|\tau^{k}(X))p_{k|k-1}^{u,k}(X)dX} & l=k.
\end{cases}\label{eq:UpdatedSingleDensities}
\end{gather}
We now factorise (\ref{eq:ThetaInitial}) as
\begin{multline}
\vartheta(X)=\rho_{k}^{u}\left[\frac{\beta_{k|k-1}^{u}(k)p^{u}(z_{k}||\mathbf{\widetilde{x}}_{k}|=1)}{\rho_{k}^{u}}p_{k|k}^{u,k}(X)\right]\\
+\rho_{k}^{u}\left[p^{u}(z_{k}|\emptyset)\sum_{l=t^{u}}^{k-1}\frac{\beta_{k|k-1}^{u}(l)}{\rho_{k}^{u}}p_{k|k}^{u,l}(X)\right]\label{eq:Rho_Form}
\end{multline}
where
\begin{equation}
\rho_{k}^{u}=\beta_{k|k-1}^{u}(k)p^{u}(z_{k}||\mathbf{\widetilde{x}}_{k}|=1)+p^{u}(z_{k}|\emptyset)\sum_{l=t^{u}}^{k-1}\beta_{k|k-1}^{u}(l).
\end{equation}
Considering $\vartheta(X)$ as (\ref{eq:Rho_Form}), this implies
that the updated existence probability and Beta parameters are
\begin{gather}
r_{k|k}^{u}=\frac{\rho_{k}^{u}r_{k|k-1}^{u}}{\rho_{k}^{u}r_{k|k-1}^{u}+p^{u}(z_{k}|\emptyset)(1-r_{k|k-1}^{u})}\\
\beta_{k|k}^{u}(l)\propto\begin{cases}
p^{u}(z_{k}|\emptyset)\beta_{k|k-1}^{u}(l) & l\in\{t^{u},...,k-1\}\\
p^{u}(z_{k}||\mathbf{\widetilde{x}}_{k}|=1)\beta_{k|k-1}^{u}(k) & l=k.
\end{cases}
\end{gather}
The updated single-trajectory densities are given by (\ref{eq:UpdatedSingleDensities}),
which finishes the proof of Lemma \ref{lem:-Given-a}.

\section{\label{subsec:GT-IEMB-Prediction-for}}

This appendix includes the prediction stage of the GT-IEMB for alive
and all trajectories.

\subsubsection*{Gaussian Densities}

We assume that target states evolve according to a linear-Gaussian
transition density $g(\cdot|x_{k-1})=\mathcal{N}(\cdot;Fx_{k-1},Q)$,
where $F$ is the state transition matrix, and $Q$ is the process
noise covariance matrix. Existing targets survive to the next time
step with a constant probability of survival $p^{S}(x)=p^{S}$, and
new targets are born according to the MB birth model (\ref{eq:MBTrajBirth})
with Gaussian single-target birth densities $\overline{x}_{b}^{l}$
and $P_{b}^{l}$.

\subsubsection*{GT-IEMB Prediction for Alive Trajectories}

Given a T-MB density of the form (\ref{eq:TrajMBDensity}) at time
step $k-1$, the prediction for the GT-IEMB is given by \cite{GarciaFernandez2020}\foreignlanguage{british}{
\begin{gather}
\overline{F}_{i}=\left[0_{1,\iota-1},1\right]\otimes F\label{eq:PredictionEq1}\\
\overline{x}_{k|k-1}^{i}=\left[(\overline{x}_{k-1|k-1}^{i})^{T},(\overline{F}_{i}x_{k-1|k-1}^{i})^{T}\right]^{T}\label{eq:PredictedTrajMean}\\
P_{k|k-1}^{i}=\left[\begin{array}{cc}
P_{k-1|k-1}^{i} & P_{k-1|k-1}^{i}\overline{F}_{i}^{T}\\
\overline{F}_{i}P_{k-1|k-1}^{i} & \overline{F}_{i}P_{k-1|k-1}^{i}\overline{F}_{i}^{T}+Q
\end{array}\right]\label{eq:PredictedTrajCovariance}\\
r_{k|k-1}^{i}=p^{S}r_{k-1|k-1}^{i}\label{eq:PredictionExistence}
\end{gather}
}where $0_{n,m}$ is an $n\times m$ matrix of zeros, and $\otimes$
denotes the Kronecker product. We also introduce birth components
in the prediction stage, from the Bernoulli birth model defined in
Section \ref{subsec:TrajDynamicModel}. The mean, covariance and birth
probabilities for each birth component is
\begin{multline}
\overline{x}_{k|k-1}^{i}=\overline{x}_{b}^{i-n_{k-1|k-1}},P_{k|k-1}^{i}=P_{b}^{i-n_{k-1|k-1}}\\
r_{k|k-1}^{i}=p_{b}^{i-n_{k-1|k-1}}\label{eq:BirthComponents}
\end{multline}
where $i\in\{n_{k-1|k-1}+1,...,n_{k-1|k-1}+n_{k}^{b}\}$.

\subsubsection*{GT-IEMB Prediction for All Trajectories}

Given a T-MB density of the form (\ref{eq:TrajMBDensity}) at time
step $k-1$, we obtain the predicted trajectory mean and covariance
of each Bernoulli component using the same prediction presented in
(\ref{eq:PredictionEq1})-(\ref{eq:PredictionExistence}) for $\overline{x}_{k-1|k-1}^{i}(l)$
and $P_{k-1|k-1}^{i}(l)$, where $l=k$. For $l\in\{t^{i},...,k-1\}$
we do not perform a prediction, meaning $\overline{x}_{k|k-1}^{i}(l)=\overline{x}_{k-1|k-1}^{i}(l)$
and $P_{k|k-1}^{i}(l)=P_{k-1|k-1}^{i}(l)$. The predicted existence
probability is given by $r_{k|k-1}^{i}=r_{k-1|k-1}^{i}$, and we also
introduce the birth components (\ref{eq:BirthComponents}).

For the set of all trajectories, we must also consider $\beta_{k|k-1}^{i}(l)$,
which is predicted using \cite{GarciaFernandez2020}
\begin{equation}
\beta_{k|k-1}^{i}(l)=\begin{cases}
\beta_{k-1|k-1}^{i}(l) & l\in\{t^{i},...,k-2\}\\
(1-p^{S})\beta_{k-1|k-1}^{i}(l) & l=k-1\\
p^{S}\beta_{k-1|k-1}^{i}(l) & l=k.
\end{cases}\label{eq:BetaPrediction}
\end{equation}
 Given that the $i$-th potential trajectory has existed at some point
and the measurements up to time step $k-1$,$\beta_{k-1|k-1}^{i}(l)$
represents the probability that the trajectory terminates at time
step $l$. When we perform the prediction, this probability does not
change for time steps $l\in\{t^{i},...,k-2\}$. The probability that
the trajectory is alive at time step $k$ is $\beta_{k-1|k-1}^{i}(k-1)$
times the probability of survival. The probability that the trajectory
terminates at time step $k-1$ is $\beta_{k-1|k-1}^{i}(k-1)$ times
the probability of death \foreignlanguage{british}{$(1-p^{S})$.}
We also clarify that while the mean and covariance $\overline{x}_{k|k-1}^{i}(l)$
and $P_{k|k-1}^{i}(l)$ for $l=k-1$ are unchanged in the prediction
stage, there is a change to $\beta_{k|k-1}^{i}(l)$ for $l=k-1$ to
account for the probability that the trajectory died at this time
step.

\section{\label{subsec:Mean-and-Covariance}}

This appendix provides the proof of Lemma \ref{lem:Moments_of_y^i}.
That is, we prove the mean and covariance of $h^{i}$, and the mean
of $R^{i}$, which are presented in (\ref{eq:Meany^i}), (\ref{eq:Covariancey^i})
and (\ref{eq:MeanR^i}), respectively.

The density of $h^{i}$ is
\begin{align}
p_{k|k-1}^{i}(h^{i}) & =\int\widetilde{f}_{k|k-1}^{i}(\widetilde{\mathbf{x}}_{k}^{i},h^{i})\delta\widetilde{\mathbf{x}}_{k}^{i}\nonumber \\
 & =(1-r_{k|k-1}^{i}(k))\delta_{0}(h^{i})\nonumber \\
 & +r_{k|k-1}^{i}(k)\int\delta_{h(x)}(h^{i})p_{k|k-1}^{i}(x)dx
\end{align}
where the mean is then given by
\begin{align}
\mathrm{E}[h^{i}] & =(1-r_{k|k-1}^{i}(k))0+r_{k|k-1}^{i}(k)\mathrm{E}^{i}[h(x)]\nonumber \\
 & =r_{k|k-1}^{i}(k)\mathrm{E}^{i}[h(x)]
\end{align}
and the covariance
\begin{align}
\mathrm{C}[h^{i}] & =\mathrm{E}[\mathrm{C}[h^{i}|\widetilde{\mathbf{x}}_{k}^{i}]]+\mathrm{C}[\mathrm{E}[h^{i}|\widetilde{\mathbf{x}}_{k}^{i}]]\nonumber \\
 & =0+\mathrm{E}\left[\mathrm{E}[h^{i}|\widetilde{\mathbf{x}}_{k}^{i}]\mathrm{E}[h^{i}|\widetilde{\mathbf{x}}_{k}^{i}]^{T}]\right]-\mathrm{E}[h^{i}]\mathrm{E}[h^{i}]^{T}
\end{align}
where
\begin{multline}
\mathrm{E}\left[\mathrm{E}[h^{i}|\widetilde{\mathbf{x}}_{k}^{i}]\mathrm{E}[h^{i}|\widetilde{\mathbf{x}}_{k}^{i}]^{T}]\right]=(1-r_{k|k-1}^{i}(k))\mathrm{E}[h^{i}|\emptyset]\mathrm{E}[h^{i}|\emptyset]^{T}+\\
r_{k|k-1}^{i}(k)\int h(x)(h(x))^{T}p_{k|k-1}^{i}(x)dx\\
=r_{k|k-1}^{i}(k)\int h(x)(h(x))^{T}p_{k|k-1}^{i}(x)dx.
\end{multline}
Therefore, the covariance of $h^{i}$ is given by
\begin{align}
\mathrm{C}[h^{i}] & =r_{k|k-1}^{i}(k)\int h(x)(h(x))^{T}p_{k|k-1}^{i}(x)dx\nonumber \\
 & -(r_{k|k-1}^{i}(k))^{2}\mathrm{E}^{i}[h(x)]\mathrm{E}^{i}[h(x)]^{T}
\end{align}
The mean of $R^{i}$ is derived analogously to the mean of $h^{i}$.

\section{\label{subsec:Conditional-Mean-and}}

This appendix proves Proposition \ref{prop:ConditionalMoments}. That
is, it provides the proof for the conditional mean and covariance
of the measurement, presented in (\ref{eq:FullConditionalMean}) and
(\ref{eq:FullConditionalCovar}), following the notation of Section
\ref{subsec:Conditional-Moments-of}.

The mean of the measurement conditioned on $\widetilde{\mathbf{x}}_{k}^{u}=\{(u,x^{u})\}$
is given by
\begin{align}
\mathrm{E}\left[z_{k}|\widetilde{\mathbf{x}}_{k}^{u}=\{(u,x^{u})\}\right] & =\int\mathrm{E}\left[z_{k}|\{(u,x^{u})\}\cup\widetilde{\mathbf{x}}_{k}^{(-u)}\right]\nonumber \\
 & \times\prod_{i=1:i\neq u}^{n_{k|k-1}}\widetilde{f}_{k|k-1}^{i}(\widetilde{\mathbf{x}}_{k}^{i})\delta\widetilde{\mathbf{x}}_{k}^{(-u)}\nonumber \\
 & =\int m(h(x^{u})+h^{(-u)})\nonumber \\
 & \times\prod_{i=1:i\neq u}^{n_{k|k-1}}\widetilde{f}_{k|k-1}^{i}(\widetilde{\mathbf{x}}_{k}^{i})\delta\widetilde{\mathbf{x}}_{k}^{(-u)}\label{eq:FullMeanMoment}
\end{align}
where we have used the result of (\ref{eq:GeneralisedMean}) and (\ref{eq:Sumh^i}).
Applying the change of variables formula for set integrals \cite[Sec. 3.5.2]{Mahler2014}
yields
\begin{align}
\mathrm{E}\left[z_{k}|\widetilde{\mathbf{x}}_{k}^{u}=\{(u,x^{u})\}\right] & =\int m(h(x^{u})+h^{(-u)})\nonumber \\
 & \times p(h^{(-u)})dh^{(-u)}
\end{align}
where $p(h^{(-u)})$ is the density of $h^{(-u)}$, whose mean $\hat{h}_{corr}^{u}$
and covariance $S_{corr}^{u}$ are given in Lemma \ref{lem:The-mean-and}.
Applying the Taylor series approximation (\ref{eq:TaylorSeriesMean})
we obtain
\begin{align}
 & \mathrm{E}\left[z_{k}|\widetilde{\mathbf{x}}_{k}^{u}=\{(u,x^{u})\}\right]\approx\nonumber \\
 & \int\left[m(h(x^{u})+\hat{h}_{corr}^{u})+M(h(x^{u})+\hat{h}_{corr}^{u})(h^{(-u)}-\hat{h}_{corr}^{u})\right]\nonumber \\
 & \times p(h^{(-u)})dh^{(-u)}\nonumber \\
 & =m(h(x^{u})+\hat{h}_{corr}^{u})\int p(h^{(-u)})dh^{(-u)}\nonumber \\
 & +M(h(x^{u})+\hat{h}_{corr}^{u})\int(h^{(-u)}-\hat{h}_{corr}^{u})p(h^{(-u)})dh^{(-u)}\nonumber \\
 & =m(h(x^{u})+\hat{h}_{corr}^{u})
\end{align}
since\foreignlanguage{british}{ $\int h^{(-u)}p(h^{(-u)})dh^{(-u)}=\hat{h}_{corr}^{u}$.}

The covariance of the measurement conditioned on $\widetilde{\mathbf{x}}_{k}^{u}=\{(u,x^{u})\}$
can be written using the law of total covariance as
\begin{align}
 & \mathrm{C}\left[z_{k}|\widetilde{\mathbf{x}}_{k}^{u}=\{(u,x^{u})\}\right]\nonumber \\
 & =\int\mathrm{C}\left[z_{k}|\{(u,x^{u})\}\cup\widetilde{\mathbf{x}}_{k}^{(-u)}\right]\times\prod_{i=1:i\neq u}^{n_{k|k-1}}\widetilde{f}_{k|k-1}^{i}(\widetilde{\mathbf{x}}_{k}^{i})\delta\widetilde{\mathbf{x}}_{k}^{(-u)}\nonumber \\
 & +\mathrm{C}\left[\mathrm{E}\left[z_{k}|\widetilde{\mathbf{x}}_{k}^{u}=\{(u,x^{u})\}\cup\widetilde{\mathbf{x}}_{k}^{(-u)}\right]|\widetilde{\mathbf{x}}_{k}^{u}=\{(u,x^{u})\}\right].\label{eq:ConditionalCovar}
\end{align}
Using (\ref{eq:GeneralisedCovariance}) and (\ref{eq:SumR^i}), the
first term in (\ref{eq:ConditionalCovar}) can be written as
\begin{align}
 & \int\mathrm{C}\left[z_{k}|\{(u,x^{u})\}\cup\widetilde{\mathbf{x}}_{k}^{(-u)}\right]\times\prod_{i=1:i\neq u}^{n_{k|k-1}}\widetilde{f}_{k|k-1}^{i}(\widetilde{\mathbf{x}}_{k}^{i})\delta\widetilde{\mathbf{x}}_{k}^{(-u)}\nonumber \\
 & =\int\Sigma(R(x^{u})+R^{(-u)})\times\prod_{i=1:i\neq u}^{n_{k|k-1}}\widetilde{f}_{k|k-1}^{i}(\widetilde{\mathbf{x}}_{k}^{i})\delta\widetilde{\mathbf{x}}_{k}^{(-u)}
\end{align}
and can be approximated using the zero-order Taylor series approximation
(\ref{eq:TaylorSeriesCovar}), which gives
\begin{align}
 & \int\Sigma(R(x^{u})+R^{(-u)})\times\prod_{i=1:i\neq u}^{n_{k|k-1}}\widetilde{f}_{k|k-1}^{i}(\widetilde{\mathbf{x}}_{k}^{i})\delta\widetilde{\mathbf{x}}_{k}^{(-u)}\nonumber \\
 & \approx\int\Sigma(R(x^{u})+R_{corr}^{u})\times\prod_{i=1:i\neq u}^{n_{k|k-1}}\widetilde{f}_{k|k-1}^{i}(\widetilde{\mathbf{x}}_{k}^{i})\delta\widetilde{\mathbf{x}}_{k}^{(-u)}\nonumber \\
 & =\Sigma(R(x^{u})+R_{corr}^{u}).
\end{align}

\noindent The second term in (\ref{eq:ConditionalCovar}) can then
be written using (\ref{eq:GeneralisedMean}) as
\begin{align}
 & \mathrm{C}\left[\mathrm{E}\left[z_{k}|\widetilde{\mathbf{x}}_{k}^{u}=\{(u,x^{u})\}\cup\widetilde{\mathbf{x}}_{k}^{(-u)}\right]|\widetilde{\mathbf{x}}_{k}^{u}=\{(u,x^{u})\}\right]\nonumber \\
 & =\mathrm{C}\left[m(h(x^{u})+h^{(-u)})|\widetilde{\mathbf{x}}_{k}^{u}=\{(u,x^{u})\}\right]
\end{align}
where we can approximate $m(h(x^{u})+h^{(-u)})$ using (\ref{eq:TaylorSeriesMean}),
and this result gives
\begin{align}
 & \mathrm{C}\left[m(h(x^{u})+h^{(-u)})|\widetilde{\mathbf{x}}_{k}^{u}=\{(u,x^{u})\}\right]\nonumber \\
 & =\mathrm{C}\left[M(h(x^{u})+\hat{h}_{corr}^{u})(h^{(-u)}-\hat{h}_{corr}^{u})|\widetilde{\mathbf{x}}_{k}^{u}=\{(u,x^{u})\}\right]\nonumber \\
 & =M(h(x^{u})+\hat{h}_{corr}^{u})S_{corr}^{u}M(h(x^{u})+\hat{h}_{corr}^{u})^{T}
\end{align}
where $m(h(x^{u})+\hat{h}_{corr}^{u})$ is omitted as its covariance
with respect to $\widetilde{\mathbf{x}}_{k}^{u}$ is 0.

For the case where $\widetilde{\mathbf{x}}_{k}^{u}=\emptyset$, the
conditional mean is now approximated by
\begin{equation}
m(h^{(-u)})\approx m(\hat{h}_{corr}^{u}).
\end{equation}
For the conditional covariance, the first term in (\ref{eq:ConditionalCovar})
reduces to
\begin{equation}
\int\mathrm{C}\left[z_{k}|\widetilde{\mathbf{x}}_{k}^{(-u)}\right]\times\prod_{i=1:i\neq u}^{n_{k|k-1}}\widetilde{f}_{k|k-1}^{i}(\widetilde{\mathbf{x}}_{k}^{i})\delta\widetilde{\mathbf{x}}_{k}^{(-u)}\approx\Sigma(R_{corr}^{u})
\end{equation}
and the second term, using the expansion
\begin{align}
m(h^{(-u)}) & \approx m(\hat{h}_{corr}^{u})+M(\hat{h}_{corr}^{u})(h^{(-u)}-\hat{h}_{corr}^{u})
\end{align}
 reduces to
\begin{align}
 & \mathrm{C}\left[m(h^{(-u)})|\widetilde{\mathbf{x}}_{k}^{u}=\{(u,x^{u})\}\right]\nonumber \\
 & =\mathrm{C}\left[M(\hat{h}_{corr}^{u})(h^{(-u)}-\hat{h}_{corr}^{u})|\widetilde{\mathbf{x}}_{k}^{u}=\{(u,x^{u})\}\right]\nonumber \\
 & =M(\hat{h}_{corr}^{u})S_{corr}^{u}M(\hat{h}_{corr}^{u})^{T}
\end{align}
where we have used a Taylor series expansion of $m(h^{(-u)})$ around
$\hat{h}_{corr}^{u}$.

\section{\label{subsec:Updating-a-Gaussian}}

This appendix details the approach used to update past trajectory
information using the current time step updated moments.

Performing the prediction stage produces a predicted single-trajectory
density, which we can view considering the current time step $k$
and all previous time steps. Following the notation in \cite{GarciaFernandez2022},
we denote the predicted mean and covariance for the current time step
as $\overline{x}=\mu_{k|k-1}^{i}$ and $P_{xx}=\Xi_{k|k-1}^{i}$,
respectively, and the predicted cross covariance is denoted $P_{xy}$.
All previous time steps have trajectory mean $\overline{y}=\bar{x}_{k|k-1}^{i}(k-1)$
and covariance $P_{yy}=P_{k|k-1}^{i}(k-1)$.

This density is Gaussian and can be written as
\begin{equation}
p(y,x)=\mathcal{N\textrm{\ensuremath{\left([y^{T},x^{T}]^{T};[\overline{y}^{T},\overline{x}^{T}]^{T},\left[\begin{array}{cc}
P_{yy} & P_{xy}^{T}\\
P_{xy} & P_{xx}
\end{array}\right]\right)}}}.
\end{equation}
The single-target IEMB update in Section \ref{subsec:GaussianUpdate}
computes the updated target mean and covariance for the current time
step, which we denote $\overline{u}_{x}=\mu_{k|k}^{i}$ and $W_{xx}=\Xi_{k|k}^{i}$,
respectively. Both are used to calculate the updated single-trajectory
density mean $\overline{u}$ and covariance $W$, which are given
by \cite{Beutler2009}
\begin{equation}
\overline{u}=\left[\overline{u}_{y}^{T},\overline{u}_{x}^{T}\right],W=\left[\begin{array}{cc}
W_{yy} & W_{xy}^{T}\\
W_{xy} & W_{xx}
\end{array}\right]
\end{equation}
where
\begin{gather}
\overline{u}_{y}=\overline{y}+G(\overline{u}_{x}-\overline{x})\\
W_{xy}=W_{xx}G^{T}\\
W_{yy}=P_{yy}-G(P_{xx}-W_{xx})G^{T}\\
G=P_{xy}^{T}P_{xx}^{-1}.
\end{gather}

For the updated trajectory mean, $\overline{u}_{y}$ corresponds to
previous trajectory states which are updated using the updated single-target
density. Under the $L$-scan approximation, this is the $L$-1 most
recent time steps. Trajectory states before $\overline{u}_{y}$ remain
unchanged, as they are outside the $L$-scan window. The covariance
matrix $W$ is also bound to size $Ln_{x}\times Ln_{x}$, which then
contains the covariance across the $L$ most recent time steps.

\section{\label{subsec:Moments-of-the}}

This appendix provides some useful results regarding the moments of
the Rician distributed measurements.

For the conditional measurement mean in (\ref{eq:RicianConditionalMean}),
we note that \cite[22.5.54]{Abramowitz1972}
\begin{equation}
L_{1/2}(x)=L_{1/2}^{(0)}(x)=M\left(-\frac{1}{2},1,x\right)
\end{equation}
where $L_{1/2}^{(0)}(x)$ is the nested Laguerre polynomial, and $M(-\frac{1}{2},1,x)$
denotes the confluent hypergeometric function of the first kind (Kummer's
function) \cite[Chapter 13]{Abramowitz1972}. To form the conditional
measurement covariance in (\ref{eq:FullConditionalCovar}), we require
the Jacobian of $m(\cdot)$, which involves partial differentiation
of $m(\cdot)$. We note that the following result proves useful \cite[13.4.8]{Abramowitz1972}
\begin{equation}
\frac{\partial}{\partial x}M(a,b,x)=\frac{a}{b}M(a+1,b+1,x).
\end{equation}

\end{document}